\documentclass{aa}  
\usepackage{graphicx}
\usepackage{txfonts}
\usepackage[colorlinks=true, linkcolor=blue, citecolor=blue, urlcolor=gray]{hyperref}
\usepackage{xcolor}
\usepackage{listings}
\usepackage{braket}
\usepackage{physics}
\usepackage[hang,small,bf]{caption}
\usepackage[subrefformat=parens]{subcaption}
\newcommand{\Msun}{{\rm  M_{\odot}}}
\newcommand{\Zsun}{Z_{\odot}}

\newcommand{\um}{\rm \mu m}

\newcommand{\Tdust}{T_{\mathrm {dust}}}

\begin{document} 

   \title{Observational signatures of the dust size evolution in isolated galaxy simulations}


   \author{Kosei Matsumoto\inst{1,2,3}
         \fnmsep \thanks{\email{kosei.matsumoto@ugent.be}}
          \and 
          Hiroyuki Hirashita \inst{4,5}
          \and
          Kentaro Nagamine\inst{5,6,7}
          \and
          Stefan van der Giessen\inst{1,8}
          \and 
          Leonard E. C. Romano\inst{9,10,11}
          \and
          Monica Relaño\inst{8,12}
          \and
          Ilse De Looze\inst{1}
          \and
          Maarten Baes\inst{1}
          \and
          Angelos Nersesian\inst{1,13}
          \and
          Peter Camps\inst{1}
          \and
          \\Kuan-chou Hou\inst{14} 
          \and          
          Yuri Oku\inst{5} 
          }

   \institute{
            Sterrenkundig Observatorium Department of Physics and Astronomy Universiteit Gent, Krijgslaan 281 S9, B-9000 Gent, Belgium\\
         \and
            Department of Physics, Graduate School of Science, The University of Tokyo, 7-3-1 Hongo, Bunkyo-ku, Tokyo 113-0033, Japan
        \and
            Institute of Space and Astronautical Science, Japan Aerospace Exploration Agency, 3-1-1 Yoshinodai, Chuo-ku, Sagamihara, Kanagawa 252-5210, Japan
        \and
            Institute of Astronomy and Astrophysics, Academia Sinica, Astronomy-Mathematics Building, No. 1, Section 4, Roosevelt Road, Taipei 10617, Taiwan
        \and
             Theoretical Astrophysics, Department of Earth and Space Science, Osaka University, 1-1 Machikaneyama, Toyonaka, Osaka 560-0043, Japan
        \and
            Kavli IPMU (WPI), The University of Tokyo, 5-1-5 Kashiwanoha, Kashiwa, Chiba 277-8583, Japan
        \and
            Department of Physics and Astronomy, University of Nevada, Las Vegas, 4505 S. Maryland Pkwy, Las Vegas, NV 89154-4002, USA
        \and
            Dept. F\'{i}sica Te\'{o}rica y del Cosmos, Universidad de Granada, Spain
        \and
            Universit\"{a}ts-Sternwarte, Fakult\"{a}t f\"{u}r Physik, Ludwig-Maximilians-Universit\"{a}t M\"{u}nchen, Scheinerstr. 1, D-81679 M\"{u}nchen, Germany
        \and
            Max-Planck-Institut f\"{u}r extraterrestrische Physik, Giessenbachstr. 1, D-85741 Garching, Germany
        \and
            Excellence Cluster ORIGINS, Boltzmannstr. 2, D-85748 Garching, Germany
        \and
            Instituto Universitario Carlos I de Física Te\'{o}rica y Computacional, Universidad de Granada, 18071, Granada, Spain
        \and
            STAR Institute, Universit\'e de Li{\`e}ge, Quartier Agora, All\'ee du six Aout 19c, B-4000 Liege, Belgium
        \and
            Physics Department, Ben-Gurion University of the Negev, Be’er-Sheva 84105, Israel
             }

   \date{Received \today}

 
  \abstract
  {}
   {We aim to provide observational signatures of the dust size evolution in the interstellar medium (ISM). In particular, we explore indicators of the polycyclic aromatic hydrocarbon (PAH) mass fraction ($q_\mathrm{PAH}$), defined as the mass fraction of PAHs relative to the total dust grains. In addition, we validate our dust evolution model by comparing the observational signatures from our simulations to observations.}
   {We used the hydrodynamic simulation code, GADGET4-OSAKA to model the dust properties of Milky Way-like and NGC 628-like galaxies representing star-forming galaxies.  This code incorporates the evolution of grain size distributions driven by dust production and interstellar processing. Furthermore, we performed post-processing dust radiative transfer calculations with SKIRT based on the hydrodynamic simulations to predict the observational properties of the simulations.}
   {We find that the intensity ratio between 8 $\um$ and 24 $\um$ ($I_\mathrm{\nu} (\mathrm{8 \, \um})/I_\mathrm{\nu} (\mathrm{24 \, \um})$) is correlated with $q_\mathrm{PAH}$ and can be used as an indicator of the PAH mass fraction. However, this ratio is influenced by the local radiation field. We suggest the 8 $\um$-to-total infrared intensity ratio ($\nu I_\mathrm{\nu} (\mathrm{8 \, \um})/ I_\mathrm{TIR}$) as another indicator of the PAH mass fraction, since it is tightly correlated with the PAH mass fraction. Furthermore, we explored the spatially resolved evolutionary properties of the PAH mass fraction in the simulated Milky Way-like galaxy using $\nu I_\mathrm{\nu} (\mathrm{8 \, \um})/ I_\mathrm{TIR}$. We find that the spatially resolved PAH mass fraction increases with metallicity at $Z\lesssim 0.2 \ \Zsun$ due to the interplay between accretion and shattering, whereas it decreases at $Z\gtrsim 0.2 \ \Zsun$ because of coagulation. Also, coagulation decreases the PAH mass fraction in regions with a high hydrogen surface density.
   Finally, we compared the above indicators in the NGC 628-like simulation with those observed in NGC 628 by \textit{Herschel}, \textit{Spitzer}, and \textit{JWST}. Consequently, we find that our simulation underestimates the PAH mass fraction throughout the entire galaxy by a factor of $\sim 8$ on average. This could be due to the efficient loss of PAHs by coagulation in our model, suggesting that our treatment of PAHs in dense regions needs to be improved.
   }
   {}


   \maketitle
   
%
\section{Introduction}\label{introduction}
Dust in the interstellar medium (ISM) plays a key role in shaping the observed properties of galaxies, in addition to indirectly affecting galaxy evolution. Dust governs various radiative processes in galaxies: dust absorbs and scatters ultraviolet (UV) and optical light from young stars, then reemiting the absorbed energy in the infrared (IR). These processes are reflected in the spectral energy distributions (SEDs) of galaxies; thus, analyses of the UV-to-IR SEDs of galaxies provide various clues to their physical properties, such as the star formation rate (SFR) and dust mass \citep[e.g.,][]{Conroy2013}. In addition, dust grains provide favorable conditions for the formation of hydrogen molecules through the grain-surface reaction; also, the shielding of UV radiation from stars by dust grains suppresses the dissociation of molecular hydrogen \citep{Hollenbach&Salpeter1971, Hollenbach&Mckee1979, Cazaux2004, Grieco2023}. Such a molecularly rich environment is favorable to the progress of star formation, with dust acting as a cooling channel in molecular clouds \citep{Ostriker1973, Vogelsberger2019}. On the contrary, photoelectric heating by dust grains is the main heating source of the photodissociation and X-ray dissociation regions in the ISM, thereby altering the ISM conditions and leading to chemical reactions \citep{Tielens&Hollenbach1985, Maloney1996, Wolfire2022}. Therefore, dust certainly stands as a crucial component of galaxy formation and evolution.

To derive dust properties in the nearby universe, many observational studies have been performed by analyzing
SEDs with IR photometric data after the appearance of \textit{Herschel} and \textit{Spitzer} telescopes, which cover wavelengths of 3--500 $\um$ \citep[e.g.,][]{daCunha2008, Leja2017, Nersesian2019, Burgarella2020}. 
Observed IR SEDs reflect the grain temperatures dependent on the grain size \citep[including the stochastic heating;][]{Draine&Anderson1985} and the material-dependent emission features at certain wavelengths.
In particular, the most prominent features seen in $\sim$3--20 $\um$ are considered to originate from PAH bands \citep[][]{Li2001, Draine&Li2007, Li2020}. Thus, the SED fitting analysis is capable of extracting the grain size distribution and the PAH abundance \citep{Galliano2008a, Galliano2021,Relano2020, Relano2022}.
\citet{Remy-Ruyer2015A&A...582A.121R} revealed an increasing trend of PAH mass fraction, which is defined as the mass fraction of PAHs relative to total dust grains, with respect to metallicity \citep[see also][]{Draine2007PAHDefinition, Aniano2020, Shim2023}. \citet{Seok2014} suggested that PAHs are produced by the shattering of carbonaceous grains, which form more efficiently via the accretion of metals on dust grains (this process is simply referred to as accretion) in regions with higher metallicity, naturally explaining the increasing trend of PAH mass fraction against metallicity.
\citet{Chastenet2019} investigated the PAH mass fraction with spatially resolved data on the Small and Large Magellanic Clouds (SMC and LMC, respectively), finding a higher PAH mass fraction in the diffuse medium in the LMC compared to that in the SMC. To explain this result, they suggested shattering large grains to PAHs is preferable over other PAH formation scenarios \citep[e.g.,  PAH formation in the envelope of AGB stars;][]{Cherchneff1992, Sloan2008, Sloan2014, Matsuura2014} and that the shattering works more efficiently in the LMC at higher metallicities.
Therefore, these observations using PAH features provide us with clues to understanding the dust processes in various ISM phases.

In the SED fitting methods, however,  the spatial resolutions of the photometric data are always reduced to the worst one, compromising the spatial dust characteristics obtained.
Meanwhile, photometric and spectroscopic observations 
that trace prominent PAH emission features are useful for deriving more detailed spatial PAH properties (mass, size, and ionization) and offering a more comprehensive understanding of dust processes in the ISM \citep{Li2020}.
Today, the James Webb Space Telescope (\textit{JWST}) has started to investigate the spatially resolved PAH features of several nearby galaxies. \citet{Egorov2023} suggested the use of the intensity ratio of the emission at 7.7 and 11.3 $\um$ to the 21 $\um$ emission ($R_\mathrm{PAH}$) for an indicator of the PAH mass fraction given by 
\begin{equation}
    R_\mathrm{PAH} = \frac{I_\nu(\mathrm{7.7 \, \um}) + I_\nu(\mathrm{11.3 \, \um})}{I_\nu(\mathrm{21 \, \um})}.
    \label{eq: R_PAH}
\end{equation}
They found that $R_\mathrm{PAH}$ decreases with the ionization fraction and implies that PAH destruction by UV radiation works in the star-forming regions.
\citet{Chastenet2023b} exhibited a decreasing trend of the 3.35 to 11.3 $\um$ intensity ratio against molecular gas fraction, implying that larger or colder grains are more abundant in denser regions. 
Although those observations imply spatially varying dust processes in the ISM, it is difficult to quantitatively infer the actual dust properties such as mass and grain sizes.
Theoretically modeling the observable signatures of dust evolution would be useful to interpret observed dust and PAH emission.

In recent decades, many studies have been conducted on dust evolution using physically motivated one-zone models \citep[e.g.,][]{Lisenfeld1998,Dwek1998,Hirashita1999}.
Essentially, such models calculate the grain size distributions for predictions of dust signatures, such as extinction curves and dust emission SEDs.
Some models describe the evolution of grain size distributions \citep[e.g.,][]{ODonnell1997,Asano2013, Hirashita2019} and have indeed been used to predict extinction curves. \citet{Hirashita2020SEDwithPAH} and \citet{Nishida2022} investigated the evolution of the dust emission SED based on the calculated grain size distribution.

Today, these evolution models of grain size distributions have been applied to hydrodynamical simulations \citep{McKinnon2018, Aoyama2017, Aoyama2020, Hou2019, Granato2021, Romano2022Dust, Narayanan2023} to calculate the evolution of dust content in galaxies in a manner consistent with the hydrodynamic evolution of the ISM.
Among them, \citet{Aoyama2020} solved the evolution of grain size distributions represented by 32 grain radius bins in an isolated galaxy simulation with a hydrodynamic simulation code, GADGET3-OSAKA. These authors found that the grain size distribution depends on the physical condition (especially the density and metallicity) of the ambient ISM. This result underlines the importance of solving the grain size distribution in a manner consistent with the hydrodynamical and chemical evolution of the ISM. \citet{Romano2022Dust} extended their models to more sophisticated subgrid models, in which the fraction of dense gas on a sub-resolution scale is varied according to the ambient density, where dust diffusion between SPH gas particles (hereafter, gas particles) is taken into account. In particular, they show that the inhomogeneity in the grain size distribution is strongly affected by the treatment of diffusion.
These simulations broadly succeed in reproducing the Milky Way extinction curve and predicting the evolution of grain size distribution, but the radiative properties are yet to be clarified based on the simulated stellar and dust distributions.

\citet{Narayanan2023} employed hydrodynamical simulations and post-processing radiative transfer calculations, which handle the grain size distribution and dust composition including PAHs, to investigate characteristics of dust emission in different types of galaxies. 
They found a significant influence of the radiation field on the ratio between global PAH luminosity and mass, inferring the importance of the radiative processes in deriving actual observational properties of dust. 
However, it has not yet been clarified how the PAH abundance can be extracted from such a strong influence of the radiation field. 
Therefore, it is crucial to investigate how grain size distributions are reflected on IR spectral features in simulations and infer observable signatures.


In this work, we model the evolution of grain size distributions in isolated galaxies with hydrodynamic simulations \citep{Romano2022Dust}
and perform radiative transfer calculations taking into account grain size distributions and different grain species such as silicate, carbonaceous dust grains, and PAHs in post-processing. Here, we assume that the main driver of PAH formation is the shattering of large carbonaceous grains, which is implied by both observational and theoretical studies \citep{Chastenet2019, Seok2014, Rau2019, Hirashita&Murga2020, Narayanan2023}.
We focus on two isolated galaxy simulations representing star-forming galaxies similar to the Milky Way and NGC 628. The Milky Way-like galaxy has been well studied in different types of simulations as a typical star-forming galaxy useful for explaining the impact of subgrid models on the galaxy evolution \citep{Kim2016, Shimizu2019, Romano2022Dust, Romano2022Mol}.
The Milky Way-like galaxy encapsulates a broad range of gas densities and metallicity, rendering it an apt experimental setup to probe the long-term dust evolution and the impact of the evolution on observable properties.
On the other hand, NGC 628 is among the most extensively studied galaxies in observations of the nearby universe, offering high spatial resolutions in the PHANGS surveys \citep{Leroy2021, Lee2023}. 
NGC 628 is one of the most nearby galaxies and a nearly face-on galaxy, allowing us to separately study dust processes in diffuse and dense ISM with minimal interference from dust extinction. Furthermore, NGC 628 is a moderately star-forming galaxy without the signature of the active galactic nucleus observed in the \textit{JWST} program 2107 \citep[PI: J. Lee;][]{Lee2023}.
Therefore, comparing the observable properties of NGC 628 between the simulation and observations serves as a robust test to validate the dust evolution model, particularly within the context of star-forming galaxies.
More comprehensive comparisons including other nearby galaxies will be presented in our future work (van der Giessen et al. in preparation).

The main goals of this paper are as follows.
First, we explore what drives the variations of grain size distributions and PAH abundances in different galaxy environments.
Second, we investigate how dust evolution is reflected in spatially resolved dust emission and SEDs at various stages of galaxy evolution.
Then, we infer and validate various indicators of PAH mass fraction
through spatially resolved comparisons with the actual PAH mass fraction in hydrodynamic simulations.
Finally, we validate our dust evolution model by spatially comparing the indicators of the PAH mass fraction between the NGC 628-like galaxy simulation and actual observations. 

This paper is organized as follows. Section~{\ref{Method}} describes the models in our hydrodynamic simulations of GADGET4-OSAKA, post-processing radiative transfer calculations by SKIRT, and the observational analysis. 
In Section~{\ref{sec: Dust evolution in isolated galaxy simulations}}, we explore the differences in the dust evolution and PAH abundance between Milky Way-like and NGC 628-like galaxies. 
In Section~{\ref{sec: Observational signatures of the dust size evolution in the Milky Way-like galaxy}}, we investigate the evolution of the spectral features in SEDs of the Milky Way-like galaxy and the indicators of PAH mass fraction using intensity ratios at various wavelengths. 
In Section~{\ref{NGC 628-like galaxy simulation}}, we perform spatially resolved comparisons for the indicators of PAH mass fraction between observations and simulations, and discuss additional physical processes required for our dust model. 
In Section~{\ref{Conclusion}}, we give our conclusions.

\section{Methods}\label{Method}
\subsection{Isolated galaxy simulations}\label{subsec: Isolated galaxy simulations}

\begin{table*}[h]
\caption{Properties of each component laid in the initial conditions of the Milky Way-like and NGC 628-like galaxy simulations}             
\label{table:hydro parameters}      
\centering                          
\begin{tabular}{l | c c |c c}        
\hline\hline                 
& Milky Way-like & & NGC 628-like & \\
\hline
Components & Mass & Particle number & Mass & Particle number\\    
\hline                        
   Gaseous disk  & $ 8.6\times10
   ^9\ \Msun$ & $1.0 \times 10^6$ & $ 3.5\times 10^{10} \ \Msun$ & $9.0 \times 10^5$\\
   Gaseous halo & $1.0 \times10^9 \ \Msun$ & $4.0 \times 10^5$ & $ 1.1 \times 10^{10} \ \Msun$ & $3.0 \times 10^5$\\
   Dark matter halo & $ 1.0\times10^{12} \ \Msun$  &  $1.0 \times 10^6$ & $1.0\times 10^{12} \  \Msun$  &  $3.0 \times 10^5$ \\
   Stellar disk & $ 3.4\times10^{10} \ \Msun$ & $1.0 \times 10^6$ & $ 2.5\times 10^{10} \ \Msun$  &  $3.0 \times 10^5$ \\
   Bulge stars & $4.3\times10^9 \ \Msun$ & $1.25 \times 10^5$ & $9.2\times 10^9 \ \Msun$  &  $3.0 \times 10^4$   \\
\hline                                   
\end{tabular}
\end{table*}

To perform isolated galaxy simulations, we used a smooth particle hydrodynamic (SPH) simulation code, GADGET4-OSAKA \citep{Romano2022Dust,Romano2022Mol}, which is a modified version of GADGET4 \citep{Springel2021}.
GADGET4-OSAKA incorporates the OSAKA feedback model along with a dust evolution model \citep{Aoyama2020,Romano2022Dust,Romano2022Mol}. The integrated feedback and dust models allow us to approach a self-consistent treatment of star formation, feedback processes, and metal and dust formation within the context of hydrodynamic evolution in galaxies. Here, we briefly summarize those models. We refer to the cited references for more details.

The star formation and feedback processes are incorporated into our code as subgrid prescriptions. Star formation occurs stochastically, guided by a star formation efficiency of $\epsilon_*=0.05$ that follows the Kennicutt-Schmidt law \citep{Kennicutt1998}. This process is initiated when gas particles 
satisfy certain conditions, specifically when $ n_\mathrm{H}> 20 \ \mathrm{cm^{-3}}$
and $T_\mathrm{gas}<10^4$ K are satisfied, where $n_\mathrm{H}$ and $T_\mathrm{gas}$ are the hydrogen number density and gas temperature, respectively.
The Osaka feedback model distinctly addresses the feedback processes of type Ia supernovae (SNe), 
type II SNe, 
asymptotic giant branch (AGB) stars, and young stars. 
The model manages energy ejections from both thermal and kinetic feedback modes \citep{Shimizu2019}.
Additionally, SNe and AGB stars eject various metal elements 
and the ejected metal mass is calculated using the CELib library \citep{Saitoh2017}.
Moreover, we utilized the GRACKLE-3 radiative cooling library \citep{Smith2017}\footnote{\url{https://grackle.readthedocs.org/}}, which solves the non-equilibrium primordial chemical network (H, He, and D), taking into account metal cooling as well as photo-heating and photo-ionization due to the UV background \citep{Haardt&Madau2012}.

The dust evolution model adopted by \citet{Romano2022Dust} is an enhanced version of models used by \citet{Hirashita2019} and \citet{Aoyama2020}. Using this model, we calculated the grain size distribution in each gas particle.
This implicitly assumes a perfect dynamic coupling between dust and gas within each gas particle.
The grains are assumed to be compact spheres with material density, $s$, such that the grain mass, $m$, and radius, $a$, are related as $m=\frac{4}{3}\pi a^3s$. The grain size distribution, denoted as $n(a)$, is defined so that $n(a)\,\mathrm{d}a$ is the number density of grains whose radii are between $a$ and $a+\mathrm{d}a$. We considered the grain radii ranging from $3.0 \times 10^{-4}$ to 10 $\mathrm{\mu}$m with 30 logarithmically spaced discrete bins.
Although we assumed a single dust species in the calculation of the grain size distribution, we later decomposed the resulting grain size distribution in various species (as described in Section \ref{subsec:SKIRT}).
This approach has the advantage of avoiding complexity in interstellar processing caused by inter-species collisions and we can still separately address the uncertainties by changing the adopted grain material properties (noted later in this work).

In our simulations, AGB stars and SNe produce large dust grains. Here, 10\% of the ejected metals from AGB stars and SNe are assumed to condense into dust grains with a log-normal grain size distribution characterized by the mean grain radius of 0.1 $\um$ and variance of 0.47 \citep{Asano2013}. Simultaneously, dust is also destroyed by sputtering in SN shocks.
Our model also takes into account a variety of interstellar dust processes including dust size growth through accretion and coagulation in the dense ISM and shattering and thermal sputtering in the diffuse ISM.
The treatment of each process in the simulation is described by \citet{Aoyama2020} and \citet{Romano2022Dust}. We refer to these papers for more details. 

We summarize the treatment of each process in the following.
Coagulation and shattering are assumed to occur by grain–grain collisions in the dense and diffuse ISM, respectively.
Both of these processes are governed by a Smoluchowski equation (or its extended version) with a Kernel function determined by the radii of colliding grains and the grain velocities determined by the coupling with the turbulent motion, which has a Kolmogorov spectrum \citep{Kolmogorov1941} and a characteristic velocity reflecting the local physical condition.
The total mass of fragments produced by shattering in each collision is determined following \citet{Kobayashi&Tanaka2010}, and is redistributed according to a power-law fragment size distribution with an index of $-3.3$ \citep{Jones1996}.
For sputtering and accretion, the grain size distribution evolves according to an advection equation in the grain radius space with an appropriate rate of grain radius increase or decrease.
The sputtering rate is taken from \citet{Tsai1995} in the hot gas, while it is regulated by the sweeping rate of SN shocks for SN destruction.
The accretion rate is determined by the collision frequency between gas-phase metals and dust.

In our hydrodynamic simulations, the parameters for the interstellar dust processes are evaluated based on the properties of silicate. In reality, however, the efficiency of interstellar dust processing should vary for different dust compositions, and the accretion efficiency of dust also varies for each gas-phase metal element \citep{Granato2021,Choban2022,Narayanan2023}.
Consequently, the timescale for grain size evolution will differ for each dust species. However, previous studies found that both silicate and carbonaceous dust trace the same evolutionary trends in the grain size distribution within a $\sim $ Gyr time lag \citep{Hou2019,Hirashita&Murga2020}.
Therefore, as far as the evolutionary path of grain size distribution on Gyr timescales is concerned, 
the treatment of the grain composition does not have a significant impact on the grain size and SED evolution over 10 Gyr time.
In particular, the final grain size distribution achieved at $t\sim 10$ Gyr, suitable for nearby galaxies in comparison, is robustly determined by the balance between shattering and coagulation, which is not sensitive to the assumed grain composition.

We note that our simulations do not have sufficient spatial resolutions for cold and dense gas ($T_\mathrm{gas}<50$ K and $n_\mathrm{H}>10^3$ cm$^{-3}$).
On the other hand, \citet{Romano2022Dust} employed a two-phase ISM subgrid model for gas particles, where it is assumed that dense and diffuse gas phases co-exist.
The mass fraction of the dense gas phase is given by
\begin{equation}
    f_\mathrm{dense} = \mathrm{min}\, \Bigg( \alpha \, \frac{n_\mathrm{H}}{1.0 \ \mathrm{cm}^{-3}}, 1.0\Bigg),\label{eq: dense gas fraction}
\end{equation}
for which we adopted $\alpha =0.12$.
Equation~(\ref{eq: dense gas fraction}) is motivated by the fact that denser gas hosts more dense clouds with molecular hydrogen \citep{Gnedin2011, Gnedin2014}. \citet{Romano2022Dust} calibrate $\alpha$ for the Milky Way-like galaxy to obtain a global dense gas fraction of $20 \ \%$, which corresponds to a mean value of the molecular gas fraction for galaxies with stellar masses of around $10^{10}$ $\Msun$ \citep{Catinella2018}.
While the temperature and density of the dense gas medium were assumed to be constant ($T_\mathrm{dense}=50$~K and $n_\mathrm{H, \ dense}=10^3$~cm$^{-3}$), we estimated the temperature and density of the diffuse gas medium in each gas particle by considering the conservation of the internal energy.
This subgrid model allowed us to treat dust processes in the diffuse and dense ISM separately according to their respective properties within each gas particle. 
Further details of the separation treatment of the two-phase ISM are described in \citet{Romano2022Dust}.

Using these models, we simulated Milky Way-like and NGC 628-like galaxies.
While the initial condition for the Milky Way-like galaxy
was taken from the AGORA project\footnote{\url{https://sites.google.com/site/santacruzcomparisonproject/}} \citep{Kim2016}, the one for the NGC 628-like galaxy was created by DICE\footnote{\url{https://bitbucket.org/vperret/dice/src/master/}} \citep{Valentin2016Dice}.
To create the initial condition of the NGC 628-like galaxy, we needed the morphological parameters of stellar and gas distributions in NGC 628. Thus, we computed those parameters based on the observationally obtained radial profiles of stars and gas (see Appendix \ref{Ap:morphological parameters of NGC628} for details). 
Moreover, we set dark matter and gaseous halos, bulge stars, and stellar and gaseous disks in the initial conditions as summarized in Table~\ref{table:hydro parameters}.

We ran both simulations until the simulation time of 10 Gyr, where the metallicity in both simulations is consistent with the observed ones \citep[][for NGC 628]{Berg2015}.
Here, we adopted an initial metallicity of $0.0001 \ \Zsun$, where $\Zsun$ is the solar metallicity \citep[$\Zsun =0.013$;][]{Asplund2009}, to start our simulations in the condition of the early universe.
For the hydrodynamic simulations, we permited the minimum smoothing length to reach $0.1\ \epsilon_\mathrm{grav}$, where $\epsilon_\mathrm{grav}$ represents the gravitational softening length. The softening lengths of the Milky Way-like and NGC 628-like galaxy simulations are 40 and 50 pc, respectively.

\subsection{Post-processing radiative transfer with SKIRT}
\label{subsec:SKIRT}

We employed the state-of-the-art three-dimensional (3D) Monte Carlo radiative transfer code, SKIRT\footnote{\url{www.skirt.ugent.be.}}
\citep{Baes2011,Camps&Baes2015,Camps2020SKIRT9} to produce the SEDs and images from UV to millimeter wavelengths.
While SKIRT is evolving into a generic radiative transfer code with capabilities to handle many radiative processes relevant to dust and gas \citep[e.g.,][]{Camps2021, Gebek2023, Bert2023, Matsumoto2023}, here we consider absorption, scattering, emission, and self-absorption only by dust. We include stochastic heating of dust grains in SKIRT \citep{Camps2015Dust}, which is particularly important for small grains and PAHs \citep{Duley1973, Purcell1976, Draine&Anderson1985}. The code is set up to deal with dust mixtures with spatially varying grain size distributions and optical properties\footnote{Configuring custom dust mixes is described in \url{https://skirt.ugent.be/root/_tutorial_custom_dust.html}}. 
SKIRT has been optimized to post-process snapshots from hydrodynamical galaxy simulations and contains mechanisms to read in SPH, adaptive mesh refinement, and moving-mesh simulations \citep{Saftly2013, Saftly2014, Camps2013, Baes&Camps2015}. 
The methodology to set up SKIRT simulations based on various types of hydrodynamical simulations is described in detail in \citet{Camps2018, Camps2022}, \citet{Kapoor2021}, \citet{Trcka2022}, and \citet{Baes2024}. In the following, we describe the details of the post-processing models of the stellar sources and dust composition used in our SKIRT simulations: 

\begin{description}
    \item[\textbf{Stellar populations:}]
    We utilized the \citet{Bruzual&Charlot2003} templates for single stellar populations with the \citet{Chabrier2003} initial mass function. The intrinsic SED of each star particle is determined based on its stellar age, metallicity, and initial stellar mass. We note that, under the initial conditions of the hydrodynamic simulations, the disk and bulge star particles are pre-assigned to make the galaxy dynamically stable. This means that some stars already exist at the simulation time of $t=0$ Gyr. Thus, we needed to construct the star formation history (SFH) before $t=0$ Gyr for the disk and bulge star particles. Since we ran the simulations until $t=10$ Gyr and assumed the galaxy age to be comparable to the cosmic age ($13.8$ Gyr), the disk and bulge stars have formed between $t=-3.8$ and $t=0$ Gyr.
    Therefore, in post-processing, we randomly assigned their ages to obtain a constant SFH in the $3.8$ Gyr interval before $t=0$ Gyr.
    The metallicity of the disk and bulge star particles is the same as the initial metallicity of $0.0001 \ \Zsun$.
    \item[\textbf{Dust models:}]
    We applied the following dust composition model to each gas particle containing dust grains in the post-processing.
    First, larger dust grains ($a > 1.3$ nm) are designated as silicate and carbonaceous dust grains.
    The grain size boundary of $a = 1.3$ nm between smaller and larger grains is taken from the definition of the maximum size of PAHs in \citet{Draine2001DustsizeDefinition} and \citet{Draine2007PAHDefinition}. 
    The mass fractions of silicate and carbonaceous dust (denoted as $f_\mathrm{silicate}$ and $f_\mathrm{carbon}$, respectively) in each gas particle are estimated based on the abundance of Si and C within the gas particle following \citet{Hirashita&Murga2020} as
\begin{eqnarray}
    f_\mathrm{silicate} &=& \frac{6 Z_\mathrm{Si}}{6Z_\mathrm{Si}+ Z_\mathrm{C}},\\
    f_\mathrm{carbon} &=& (1-f_\mathrm{silicate}),
\end{eqnarray}
    where $Z_\mathrm{Si}$ and $Z_\mathrm{C}$ represent the mass abundances of Si and C within each gas particle, respectively. The factor of 6 is obtained from the mass fraction of Si in silicate, as described in \citet{Hirashita&Kuo2011} and \citet{Hirashita&Murga2020}.
    As shown by \citet{Hirashita2020SEDwithPAH}, our model tends to underpredict PAH emission \citep[see also][]{Chang2022}. To improve our predictions in a maximally acceptable manner within our model framework,
    we assume that all the smaller grains ($a \leq 1.3$ nm) are carbonaceous. This is consistent with the grain model obtained by \citet{Weingartner2001} as a result of fitting to the Milky Way extinction curve: in their model, the grain size distribution of silicate has a steeper decline toward small grains while that of carbonaceous dust is enhanced at small radii.
    As a consequence of this assumption, the grains with $a\leq 1.3$ nm are composed of PAH$^0$, PAH$^+$, and non-aromatic carbonaceous grains. Since the aromatization and its inverse reaction predominantly occur in the diffuse and dense media, respectively, the mass fractions of aromatic (PAH$^0$ and PAH$^+$) carbonaceous grains correspond to the diffuse gas fraction ($f_\mathrm{aroma}= 1-f_\mathrm{dense}$; \citealt{Hirashita&Murga2020}).
    In the analysis of this paper, the PAH mass fraction is given by 
\begin{equation}
       q_\mathrm{PAH}=\frac{M_\mathrm{PAH \ (a\leq 1.3 \ \mathrm{nm})}}{M_\mathrm{(total \ dust)}} \times 100. 
\end{equation}
    
    Finally, to estimate the mass fraction for the PAH$^0$ and PAH$^+$ in the aromatic carbonaceous dust grains of each gas particle, we use the ionization fraction
    from the ``standard'' ionization model in \citet{Draine2021} and \citet{Hensley2023}:
\begin{eqnarray}    
    f_\mathrm{ion}(a) = 1-\frac{1}{1+a/10 \, \AA }.
\end{eqnarray}
    The opacities and calorimetric properties of PAHs, silicate, and carbonaceous grains (assumed to be graphite) follow the prescription of \citet{Draine&Li2007}.
\item[\textbf{Instrument setups of the Milly Way-like galaxy:}]
We observed the galaxy at a distance of 10 Mpc and inclination angles of 0 and 90 $^\circ$. For generating the synthetic observational data, we set the fields of view to ($40$ kpc)$^2$ and adopt a pixel size of 50 pc, which is close to the gravitational softening length of the Milky Way-like galaxy simulation.
We selected pixels from each band data with an intensity ($I_\nu>0.01$ $\mathrm{MJy\,sr^{-1}}$) to exclude those with significant Monte Carlo noise in SKIRT.

\item[\textbf{Instrument setups of the NGC 628-like galaxy:}]
NGC 628 is located at a distance of 9.77 Mpc, and the inclination angle is 8.9 $^\circ$ \citep{Leroy2021}. Thus, the simulated galaxy is observed under the same conditions as the actual galaxy. To compare this simulation with the actual observations in Section \ref{NGC 628-like galaxy simulation}, we used the field of view of ($20$ kpc)$^2$ and a pixel size of 50 pc, which corresponds to the gravitational softening length of the NGC 628-like galaxy simulation. 
The pixels of the data in each band with an intensity ($I_\nu>0.01$ $\mathrm{MJy\,sr^{-1}}$) are selected.
When we compared the simulation data with the \textit{Herschel} and \textit{Spitzer} data, the resolution was reduced to 600 pc to make it consistent with the beam size of the {\textit{Herschel}} PACS $160$ $\um$ for NGC 628 (see also Section \ref{subsec: Observational data and analysis}). 

\end{description}

\subsection{Observational data and analysis}\label{subsec: Observational data and analysis}

For the observational photometric data of NGC 628, we used the Dustpedia database \citep{Davies2017, Clark2018}\footnote{\url{http://dustpedia.astro.noa.gr}}, Spitzer Local Volume Legacy survey \citep{Dale2009}, and PHANGS (Physics at High Angular resolution in Nearby GalaxieS) survey \citep{Lee2023}\footnote{\url{https://sites.google.com/view/phangs/home/data}} for \textit{Herschel}, \textit{Spitzer}, and \textit{JWST}, respectively. More specifically, we used the {\textit{Herschel}} PACS $70$ and $160~\um$ bands, the {\textit{Spitzer}} IRAC $8~\um$ and MIPS $24~\um$ bands, and {\textit{JWST}} MIRI $7.7$, $11.3$, and $21~\um$ bands.
After we obtained the data, we convolved each image with the kernel function of the point spread function of each instrument. 
The kernel functions were taken from the Astronomical Convolution-kernel repository \citep{Aniano2011}\footnote{\url{https://www.astro.princeton.edu/~draine/Kernels.html}}.
To accomplish the comparison between the observations and the NGC 628-like galaxy simulation in Section \ref{NGC 628-like galaxy simulation}, we averaged the observational data so that the spatial resolutions of \textit{JWST} images are reduced to 50 pc, which corresponds to the softening length of the simulation.
The resolutions of the \textit{Herschel} and \textit{Spitzer} images were unified to 600 pc, corresponding to the beam size of the {\textit{Herschel}} PACS $160$ $\um$.
After these processes, we only adopted the pixels whose intensity is above the 1 sigma level of the background noise in the original photometric data.

\section{Overall evolution of the galaxy properties of the isolated galaxy simulations}\label{sec: Dust evolution in isolated galaxy simulations}
\begin{figure}[]
    \includegraphics[width=0.48\textwidth]{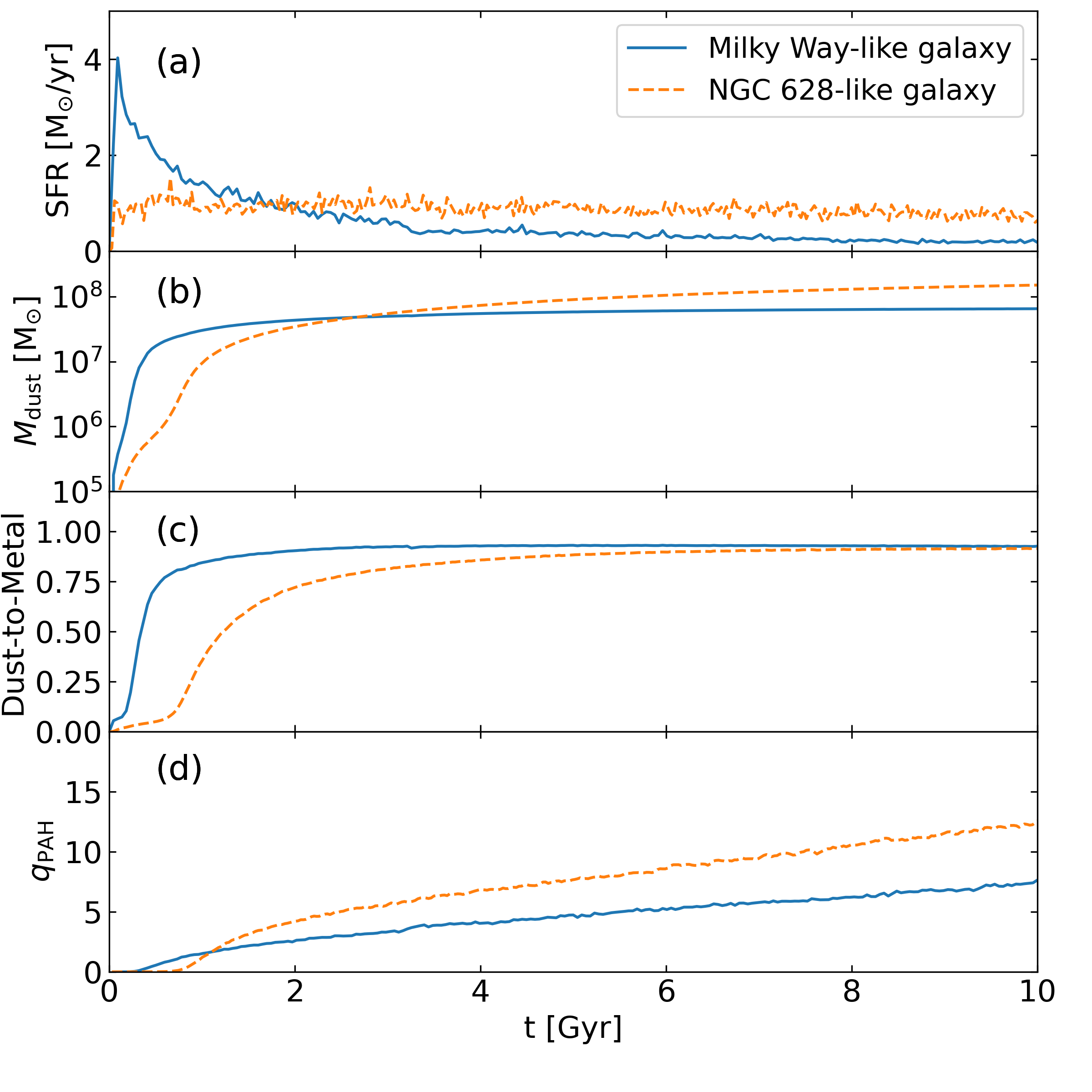}
    \caption{Time evolution of global physical properties of the Milky Way-like and NGC 628-like galaxies (blue solid and orange dashed lines, respectively). The evolution of (a) SFR, (b) dust mass, (c) dust-to-metal mass ratio, and (d) PAH mass fraction are shown.}%
    \label{fig: Dust evolutions of isolated galaxy models}
\end{figure}
Here, we provide overviews of the simulated galaxies and show the time evolution of global quantities that are important to understand the dust evolution for the Milky Way-like and NGC 628-like cases.

Figure \ref{fig: Dust evolutions of isolated galaxy models}(a) shows the SFH of the Milky Way-like and NGC 628-like galaxies.
In the Milky Way-like galaxy, a large amount of gas assigned in the galactic center starts to contract, leading to explosive star formation in the early stages of the simulation (hereafter, referred to as the starburst phase). 
Subsequently, star formation and its feedback gradually deplete the gas in the center of the galaxy, resulting in a decrease in the SFR.
In contrast, star formation occurs moderately in the NGC 628-like galaxy. The radial gas profile of the NGC 628-like galaxy is flatter than that of the Milky Way-like galaxy (See Appendix~\ref{Ap:morphological parameters of NGC628}) and the gas in the entire NGC 628-like galaxy is not dense enough to cause explosive star formation. Thus, the gas is slowly consumed by star formation with $\mathrm{SFR} \sim 1.0$ $\mathrm{M_{\odot}~yr^{-1}}$ over 10 Gyr.

The different SFHs between the two galaxies lead to their different dust evolution.
The total dust mass increases more rapidly for the Milky Way-like galaxy at $t<1.0$ Gyr than for the NGC 628-like galaxy, as shown in Fig.~\ref{fig: Dust evolutions of isolated galaxy models}(b) since more dust is produced in the starburst phase of the former galaxy.

The dust-to-metal mass ratio gives us insights into the efficiency of dust accretion in the ISM since dust grains grow by acquiring gas-phase metals \citep[e.g.,][]{Hirashita1999, Draine2009}.
Figure \ref{fig: Dust evolutions of isolated galaxy models}(c) shows that the dust-to-metal mass ratio of the Milky Way-like galaxy increases more steeply than that of the NGC 628-like galaxy since the quicker metal enrichment in the starburst phase of the Milky Way-like galaxy leads to more efficient accretion of gas-phase metals onto dust grains.

\begin{figure}[]
    \includegraphics[width=0.5\textwidth]{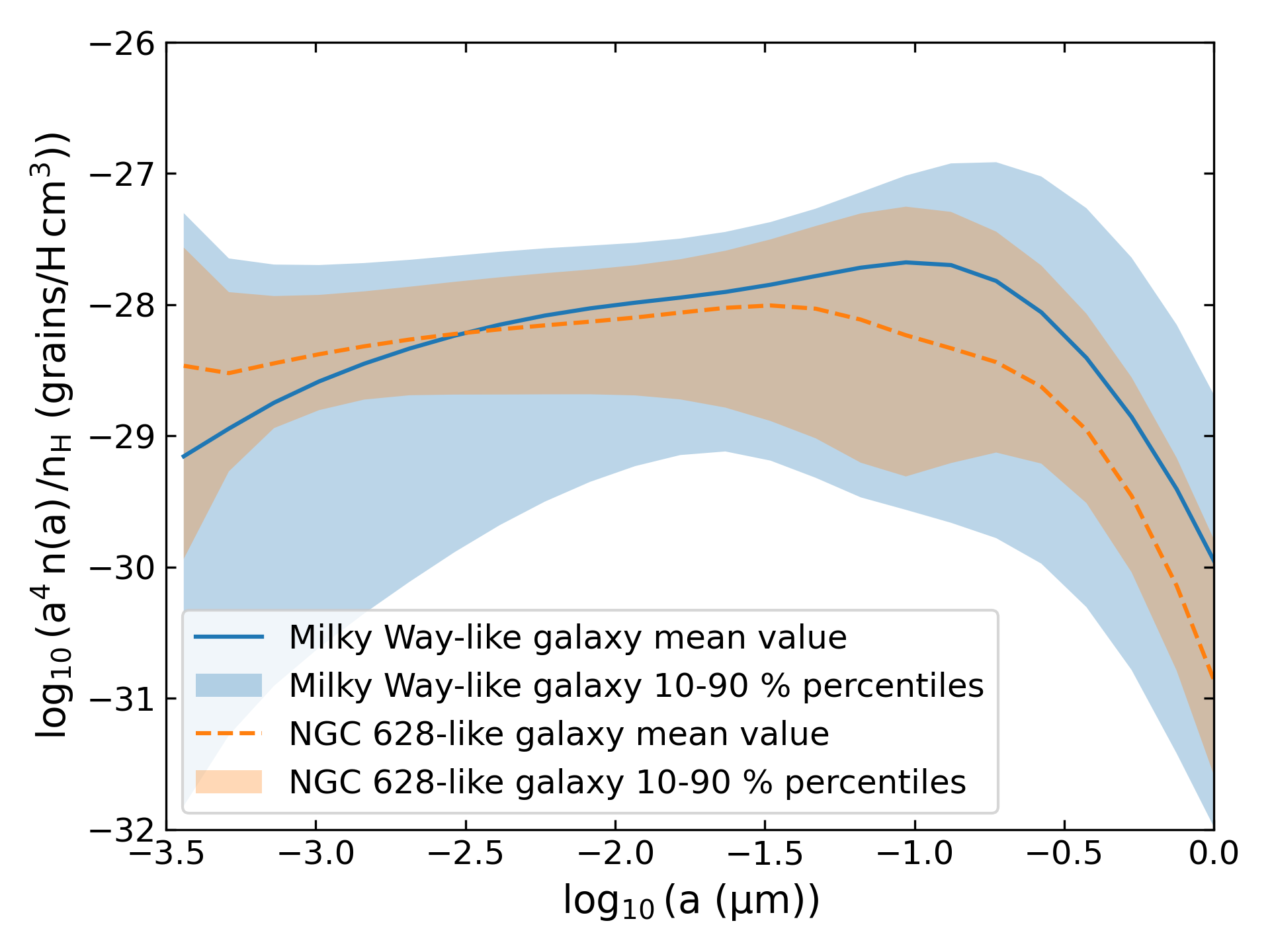}
    \caption{Grain size distribution in the Milky Way-like (blue solid line) and NGC 628-like (orange dashed line) galaxies at $t=10$ Gyr. The shades in the same color show the range of 10th--90th percentiles for the sample of gas particles.}%
    \label{fig: Comparison of the dust size distribution of the Milky-way like and NGC 628-like galaxies}
\end{figure}

Figure \ref{fig: Dust evolutions of isolated galaxy models}(d) shows the evolution of the PAH mass fractions, which indicates the efficiency of shattering since PAHs represent small grains.
The PAH mass fraction in the Milky Way-like galaxy increases less steeply than the NGC 628-like one, since the former contains more amounts of dense gas, and hence, shattering is less efficient.
The difference in the ISM conditions between these galaxies also affects their grain size distributions at $t=10$ Gyr, as shown in Fig.~\ref{fig: Comparison of the dust size distribution of the Milky-way like and NGC 628-like galaxies}.
The mean grain size distributions have different slopes between the two simulations in such a way that the Milky Way-like galaxy (blue solid line) is more dominated by large grains than the NGC 628-like galaxy (orange dashed line). 
At $t=10$ Gyr, the shape is determined by the balance between coagulation and shattering.
Coagulation turns more small grains into large grains in the Milky Way-like galaxy, since the galaxy contains more amounts of dense gas.
These results suggest that different ISM conditions and SFHs cause different dust evolution. 

\section{Observational signatures of the grain size evolution in the Milky Way-like galaxy}\label{sec: Observational signatures of the dust size evolution in the Milky Way-like galaxy}
To investigate observational signatures of the grain size evolution, we focus on the results from the Milky Way-like galaxy simulations.
The Milky Way-like galaxy simulation from the AGORA project captures various ISM environments with a broad range of gas densities and metallicities and can track the dust evolution over long periods of 10 Gyr.
We note that the following discussion in this section does not significantly change even if we involve the results of the NGC 628-like galaxy simulation.
Also, to scrutinize the ISM processes across the whole galaxy with minimal interference from dust extinction, we use the synthetic observation data of a face-on view of the Milky Way-like galaxy in Section \ref{subsec: Indicator of the PAH mass fraction}. 

\subsection{Influences of the dust size evolution on dust emission}\label{subsec: Dust emission in the Milky-way like galaxy}
\begin{figure}[h]
    \includegraphics[width=0.48\textwidth]{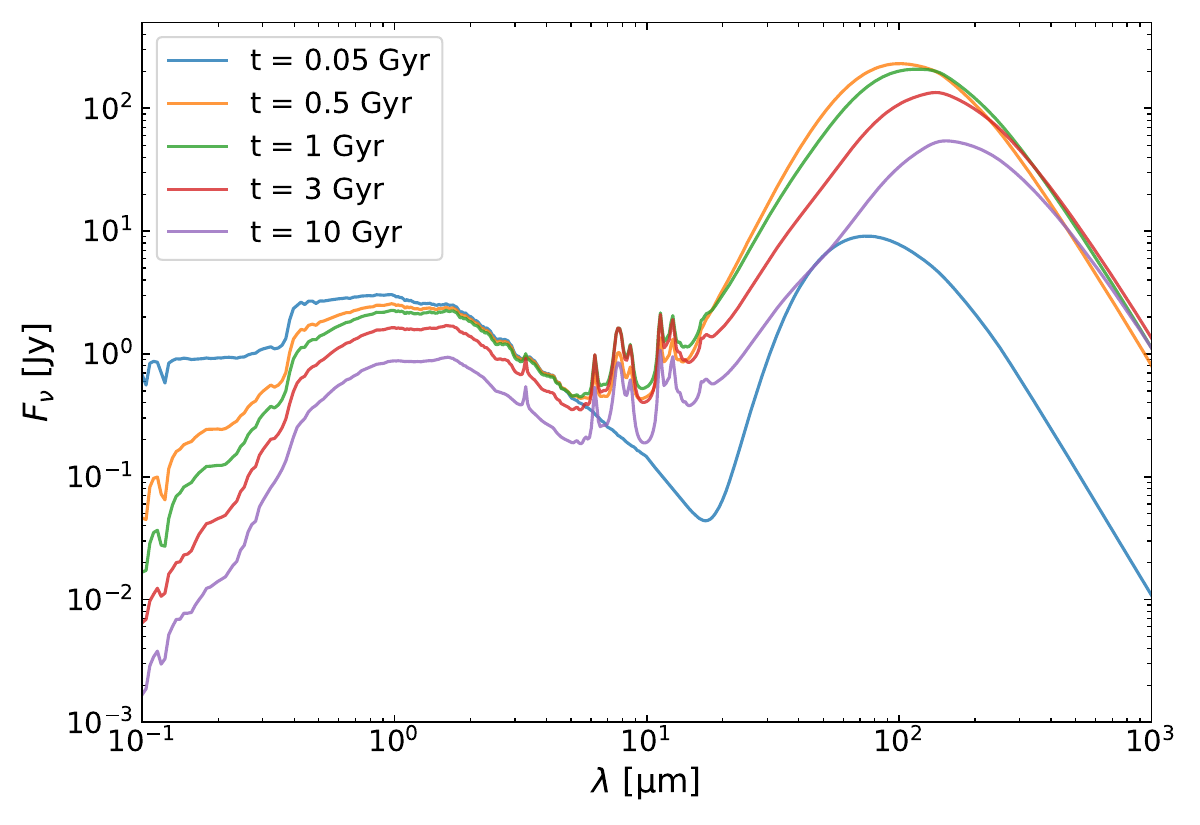}
    \caption{Evolution of the global SED for the Milky Way-like galaxy simulation. The lines in different colors correspond to the total SEDs at $t=0.05, \ 0.5, \ 1, \ 3,$ and $10$ Gyr (cyan, orange, green, red, and purple lines, respectively). The observed flux density is estimated at a distance of 10 Mpc.}%
    \label{fig: SED evolution of the MW-like galaxy}
\end{figure}

\begin{figure*}        
  \includegraphics[width=1.0\textwidth]{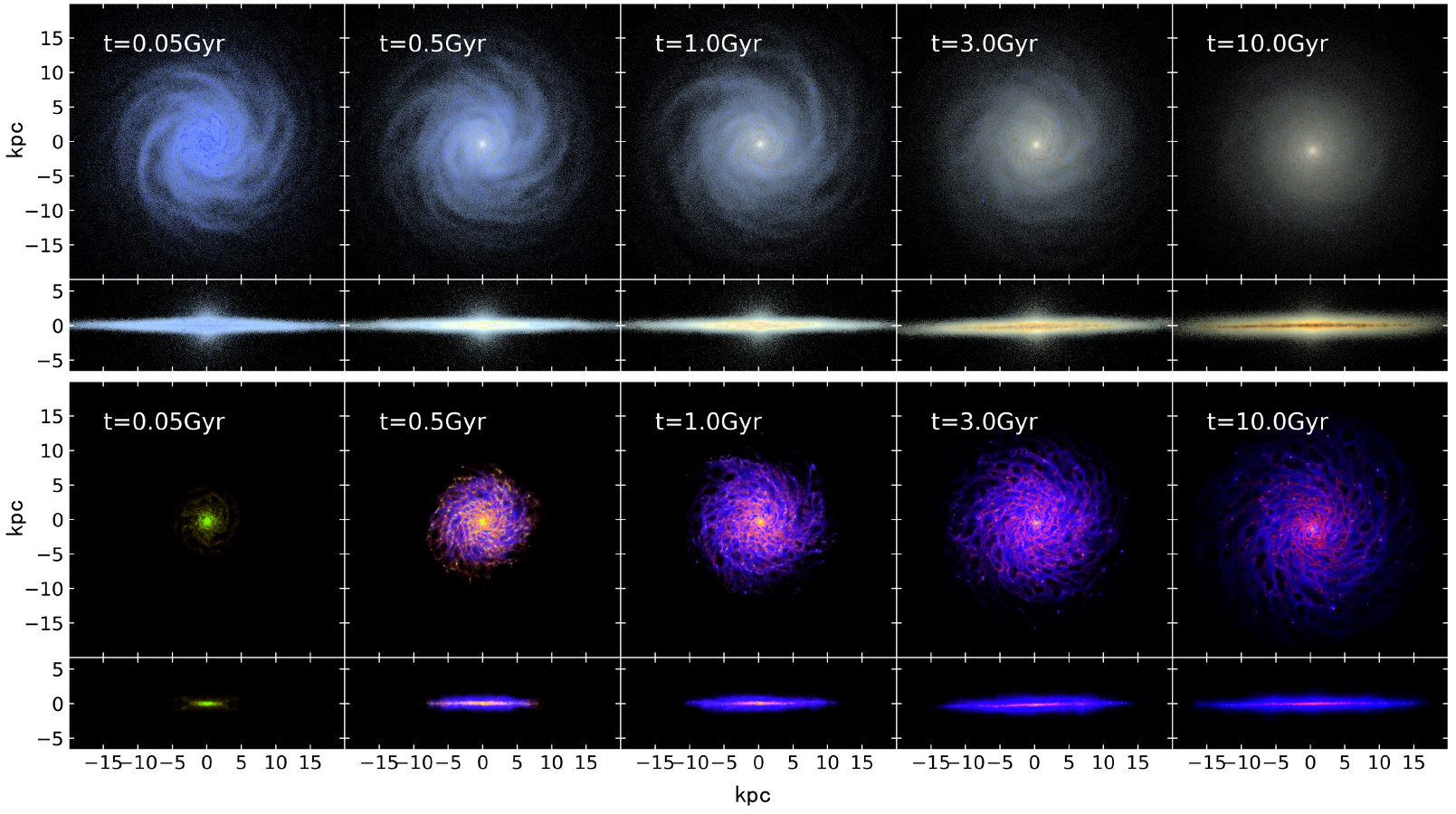}
     \caption{Evolution of stellar (first and second rows) and dust emission maps (third and fourth rows) of the Milky Way-like galaxy simulation. The panels show the emission maps at $t=0.05, \ 0.5, \ 1, \ 3,$ and $10$ Gyr from left to right.
     The images on the first and third (second and fourth) rows are face-on (edge-on) views.
     The stellar emission maps consist of the SDSS $i$ (red), $r$ (green), and $g$ (blue) fluxes with an additionally enhanced blue level for the \textit{GALEX} FUV flux. The dust emission maps are synthesized by the {\textit{Spitzer}} IRAC $8~\um$ (blue), {\textit{Herschel}} PACS $70~\um$ (green), and {\textit{Herschel}} PACS $160~\um$ (red) fluxes.}%
     \label{fig: Evolution of band-maps of the MW-like galaxy simulation}
\end{figure*}
We investigate how dust evolution is reflected in SEDs at various stages of galaxy evolution.
Figure \ref{fig: SED evolution of the MW-like galaxy} shows the total SEDs of the Milky Way-like galaxy simulation at $t=0.05, \ 0.5, \ 1, \ 3$, and $10$ Gyr.
At $t=0.05$ Gyr, only emission of large dust grains can be observed around $\lambda=70$ $\um$, as dust production is dominated by stellar sources, which predominantly produce large grains. 
At $t \geq 0.5$ Gyr, the emission of small dust grains and PAHs becomes noticeable around $\lambda =10~\um$.
This is due to shattering, which acts as a source of small grains, and accretion, which increases the abundance of small grains \citep[see also][]{Asano2013, Hirashita2019}.
Given the monotonic decrease in the SFR in the Milky Way-like galaxy at later epochs, the intrinsic radiation from young stars at UV to optical wavelengths is weakened with time.
Consequently, the peak of dust emission in the SED shifts from  $\lambda=70$ to $150$ $\um$ over time. This shift corresponds to a decrease in the dust temperature from $\Tdust=40$ to $20$ K. 
On the other hand, the dust mass monotonically increases with time mainly through the accretion of gas-phase metals. Thus, at $t=1$ Gyr, the total IR luminosity reaches its maximum.

Figure \ref{fig: Evolution of band-maps of the MW-like galaxy simulation} shows the stellar and dust emission maps from the Milky Way-like galaxy simulation at simulation times of $t=0.05, \ 0.5, \ 1.0, \ 3.0,$ and $10$ Gyr.
The first two rows show the stellar emission seen from face-on and edge-on with colors synthesized using the SDSS $i$ (red), $r$ (green), and $g$ (blue) fluxes and the \textit{GALEX} FUV flux (additionally enhanced blue color).
As viewed face-on, old stars radiate throughout the entire galaxy (appearing in white), while UV radiation from young stars (blue) originates from star-forming regions, which tend to be concentrated in dense clumps along the spiral arms.
As viewed face-on and edge-on, the effect of dust attenuation is prominent at later stages of the galaxy evolution when dust formation has proceeded enough to shield the stellar radiation. Spiral-shaped and horizontally extended dark lanes in the stellar emission develop from face-on and edge-on views, respectively.

The bottom two rows of Fig.~{\ref{fig: Evolution of band-maps of the MW-like galaxy simulation}} show the dust emission maps viewed face-on and edge-on consisting of {\textit{Spitzer}} IRAC $8~\um$ (red), {\textit{Herschel}} PACS $70~\um$ (green), and {\textit{Herschel}} PACS $160~\um$ (blue) fluxes.
The IRAC ${8~\um}$ band mainly traces the emission from PAHs heated by radiation from young stars, while the PACS ${70~\um}$ and ${160~\um}$ bands trace emission from warm and cold large dust grains, respectively.
At $t=0.05$ Gyr, large grains are produced by stars in the central region of the galaxy, and the emission from the warm large grains heated by young stars is prominent. 
At $t \geq 0.5$ Gyr, stars produce large dust grains in the spiral arms, and simultaneously accretion and shattering become prevalent in the dense and diffuse ISM, respectively. These processes increase small grains and PAHs, leading to significant emission in the MIR (i.e., the color becomes bluer in the images).
At $t \geq 1$ Gyr, coagulation also becomes efficient in the dense ISM, enhancing the emission from large grains in the central regions and spiral arms of the galaxy. The emission of small grains and PAHs is prominent in diffuse regions along the spiral arms.
At later times, as star formation activity declines, the overall dust emission becomes weaker, and the color becomes redder because of lower dust temperature.
Finally, at $t=10$ Gyr, the functional shape of the grain size distribution in each region converges to the one determined by the balance between the shattering and coagulation rates (see also Fig.~\ref{fig: Comparison of the dust size distribution of the Milky-way like and NGC 628-like galaxies}). Since the overall density is higher in the central regions (thus, coagulation is more enhanced), the emission from large and small grains is relatively prominent in the inner and outer regions of the galaxy, respectively.






\subsection{Indicators of the PAH mass fraction} \label{subsec: Indicator of the PAH mass fraction}
We investigate promising indicators of PAH mass fractions in the ISM using multi-wavelength intensity ratios in the SKIRT simulation of the Milky Way-like galaxy. 
The validity of the adopted ratios as a good indicator is tested against actual PAH mass fractions of the Milky Way-like galaxy simulation.
Since the PAH emission is the most prominent in the 8 $\um$ band among the bands available for \textit{Spitzer} and \textit{JWST}, we construct the indicators by combining the 8 $\um$ band and other wavelengths.
We focus on the following two: one is
the intensity ratio using representative MIR bands, that is, the 8 to 24 $\um$ intensity ratio, $I_\mathrm{\nu} (\mathrm{8 \, \um})/I_\mathrm{\nu} (\mathrm{24 \, \um})$,  and the other is the $8$ $\um$ to the total infrared intensity ratio, $\nu I_\mathrm{\nu} (\mathrm{8 \, \um})/ I_\mathrm{TIR}$. Here, we denote the intensity per frequency at wavelength $\lambda$ as $I_\nu (\lambda)$.
Also, we note that we investigate those indicators with the spatially resolved information of the 50-pc pixel data constructed in Section \ref{subsec:SKIRT}.
The following discussion focuses on the Milky Way-like galaxy but does not significantly change for the NGC 628-like galaxy simulation. We discuss the difference of the PAH indicator, $\nu I_\mathrm{\nu} (\mathrm{8 \, \um})/ I_\mathrm{TIR}$, between the Milky Way-like and NGC 628-like galaxy simulations in Appendix \ref{Ap: Indicators of the PAH mass fraction of the NGC 628-like galaxy}.

\subsubsection{$I_\mathrm{\nu} (\mathrm{8 \, \um})/I_\mathrm{\nu} (\mathrm{24 \, \um})$} \label{subsubsec: Signatures of dust size evolution I8/I24}

\begin{figure}
\includegraphics[width=0.5\textwidth]{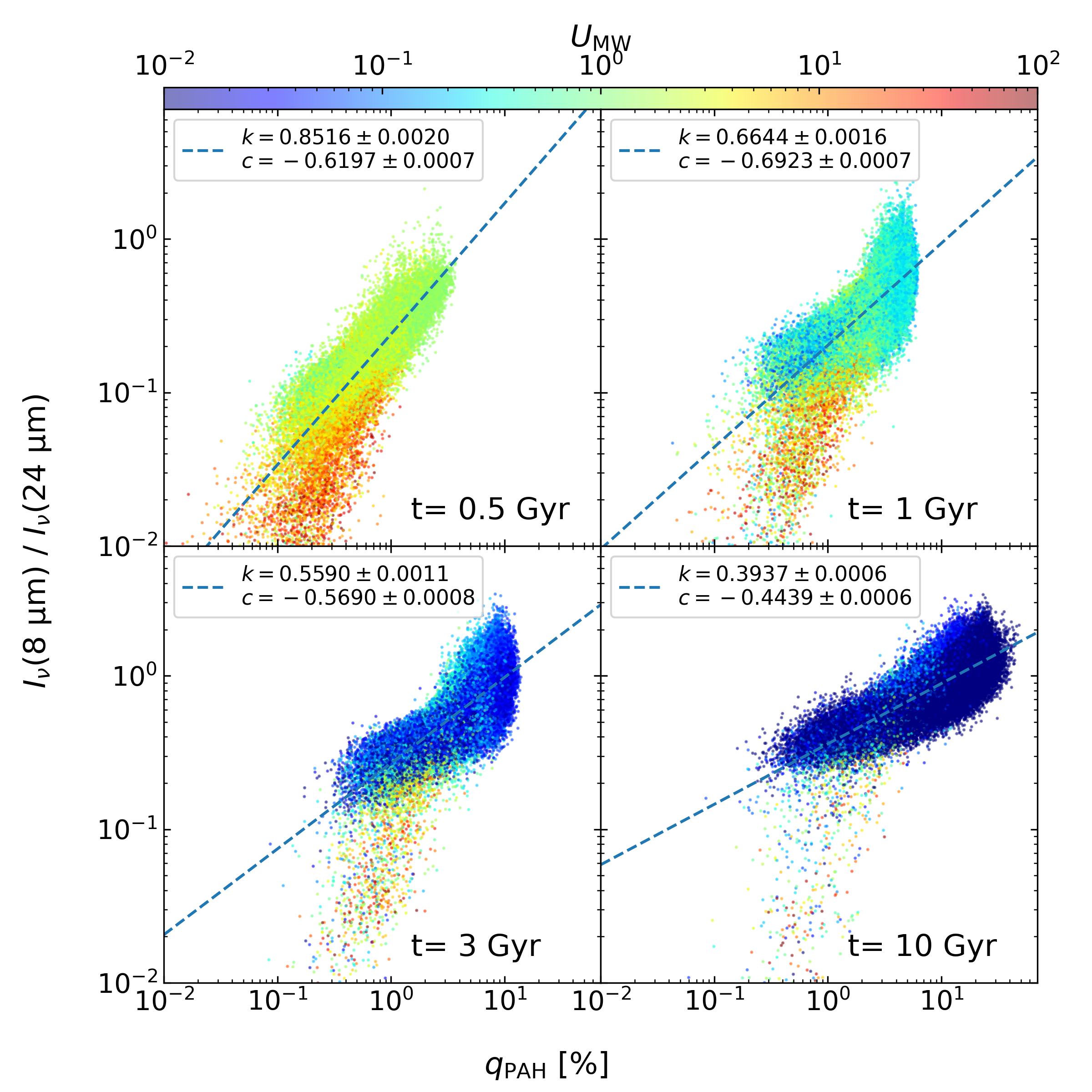}
     \caption{Pixel-based relation between the 8-to-24 $\um$ intensity ratio and the PAH mass fraction of the Milky Way-like galaxy simulation at $t=0.5$, 1.0, 3.0, and 10 Gyr from top left to bottom right.  The color represents the density-weighted mean intensity at 1000 \AA\ relative to the Milky Way value as shown by the color bar on the top. Blue dashed lines indicate the best-fit relation given by Eq.~(\ref{eq: linear regression}), and the legends show the values of the coefficients $k$ and $c$.}
     \label{fig: 8-24 ratio PAH mass fraction}
\end{figure}
\begin{figure}
\includegraphics[width=0.5\textwidth]{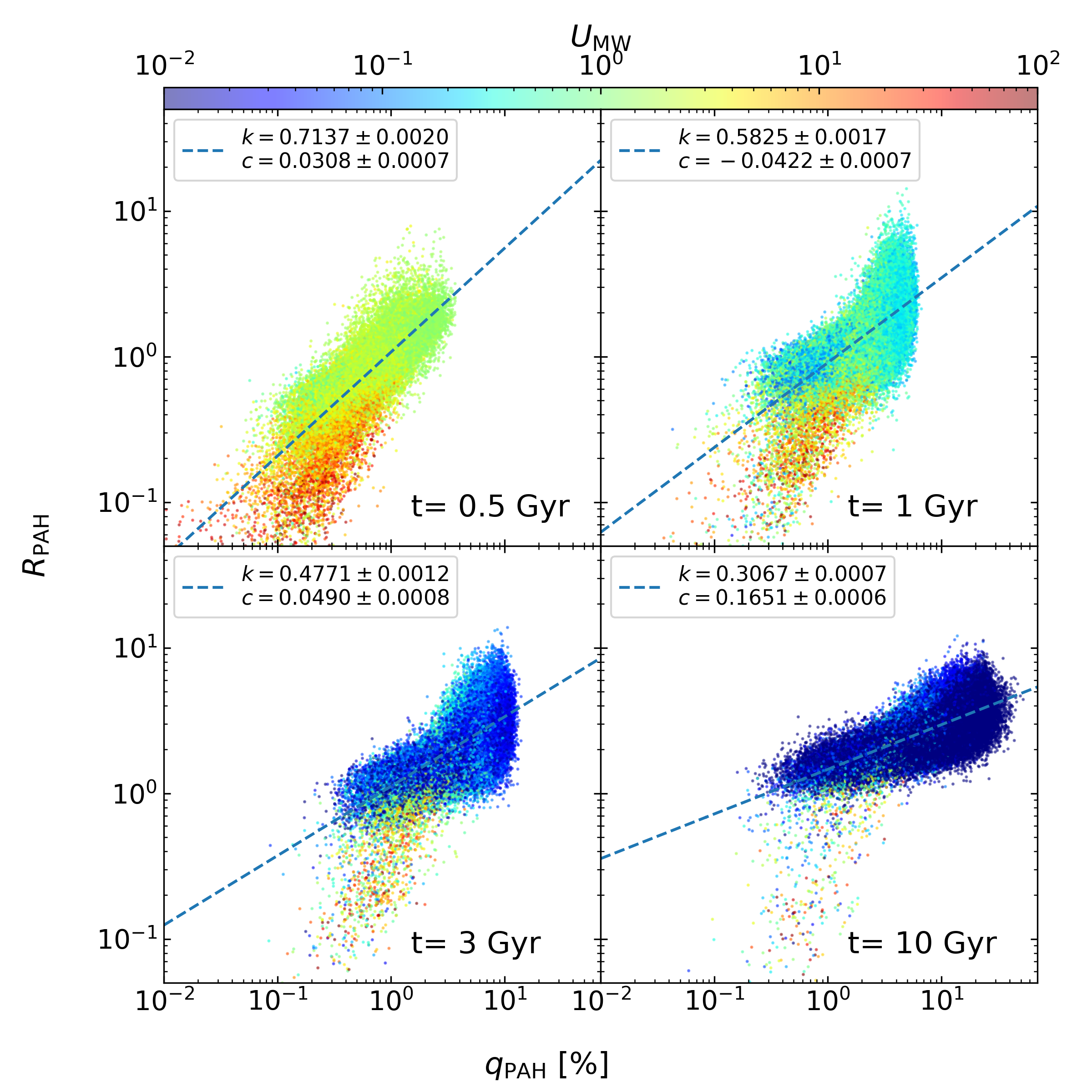}
     \caption{Same as Fig. \ref{fig: 8-24 ratio PAH mass fraction} but for $R_\mathrm{PAH}$.}
     \label{fig: RPAH PAH mass fraction}
\end{figure}
Figure \ref{fig: 8-24 ratio PAH mass fraction} shows the pixel-based relation between PAH mass fraction of $q_\mathrm{PAH}$ and $I_\mathrm{\nu} (\mathrm{8 \, \um})/I_\mathrm{\nu} (\mathrm{24 \, \um})$ at $t= 0.5, \ 1.0, \ 3.0,$ and $10$~Gyr. 
To investigate the effect of the radiation field (or dust heating), we color-code the data points with the density-weighted mean intensity at 1000~\AA\ relative to the Milky Way value of $10^6$ photons $\mathrm{cm^{-2}~s^{-1}~eV^{-1}}$, $U_\mathrm{MW}$.
We fit the data at each time by the following equation:
\begin{equation}
    \mathrm{log}_{10} (\mathrm{Intensity \ Ratio}) = k \, \mathrm{log}_{10}(q_\mathrm{PAH})+c, \label{eq: linear regression}
\end{equation}
where $k$ and $c$ are coefficients derived from the fitting, and $k$ is associated with the radiative processes of dust grains.
The best-fit relation at each time is indicated as blue dashed lines in Fig.~\ref{fig: 8-24 ratio PAH mass fraction}.
We find that $I_\mathrm{\nu} (\mathrm{8 \, \um})/I_\mathrm{\nu} (\mathrm{24 \, \um})$ correlates with $q_\mathrm{PAH}$ at each time. The 8 $\um$ band is almost purely dominated by the PAH emission, while the $24$ $\um$ band is contaminated by emission from larger dust grains with $a \gtrsim 20$ \AA\ \citep{Draine&Li2007}, so that $I_\mathrm{\nu} (\mathrm{8 \, \um})/I_\mathrm{\nu} (\mathrm{24 \, \um})$ increases with $q_\mathrm{PAH}$.
However, we also find that stronger radiation fields decrease the ratio at a fixed $q_\mathrm{PAH}$.
This is because stronger radiation fields cause higher dust temperatures, shifting the large grain emission toward shorter wavelengths and raising emission at 
24 $\um$ \citep{Draine&Li2007,Schreiber2018}.
Consequently, given the monotonic decrease in SFR with time in the Milky Way-like galaxy, the best-fitting slope becomes shallower with time.
The above dependence on the radiation field, however, is not strong enough to break the correlation between
$I_\mathrm{\nu} (\mathrm{8 \, \um})/I_\mathrm{\nu} (\mathrm{24 \, \um})$ and $q_\mathrm{PAH}$ at each time.
Therefore, $I_\mathrm{\nu} (\mathrm{8 \, \um})/I_\mathrm{\nu} (\mathrm{24 \, \um})$ can be an indicator of the PAH mass fraction with a tendency of underestimating the PAH mass fraction in regions of strong radiation fields, such as star-forming regions.
In particular, $I_\mathrm{\nu} (\mathrm{8 \, \um})/I_\mathrm{\nu} (\mathrm{24 \, \um})$ has the advantage of using only two bands in the MIR, which is useful if the available wavelength coverage is limited.
We note that the above results are also applicable to an indicator of PAH mass fraction, $R_\mathrm{PAH}$, for \textit{JWST} as shown in Fig. \ref{fig: RPAH PAH mass fraction}.

\subsubsection{$\nu I_\nu (\mathrm{8 \, \um})/ I_\mathrm{TIR}$}\label{subsubsec: Signatures of dust size evolution I8/ITIR}
\begin{figure}
        \includegraphics[width=0.5\textwidth]{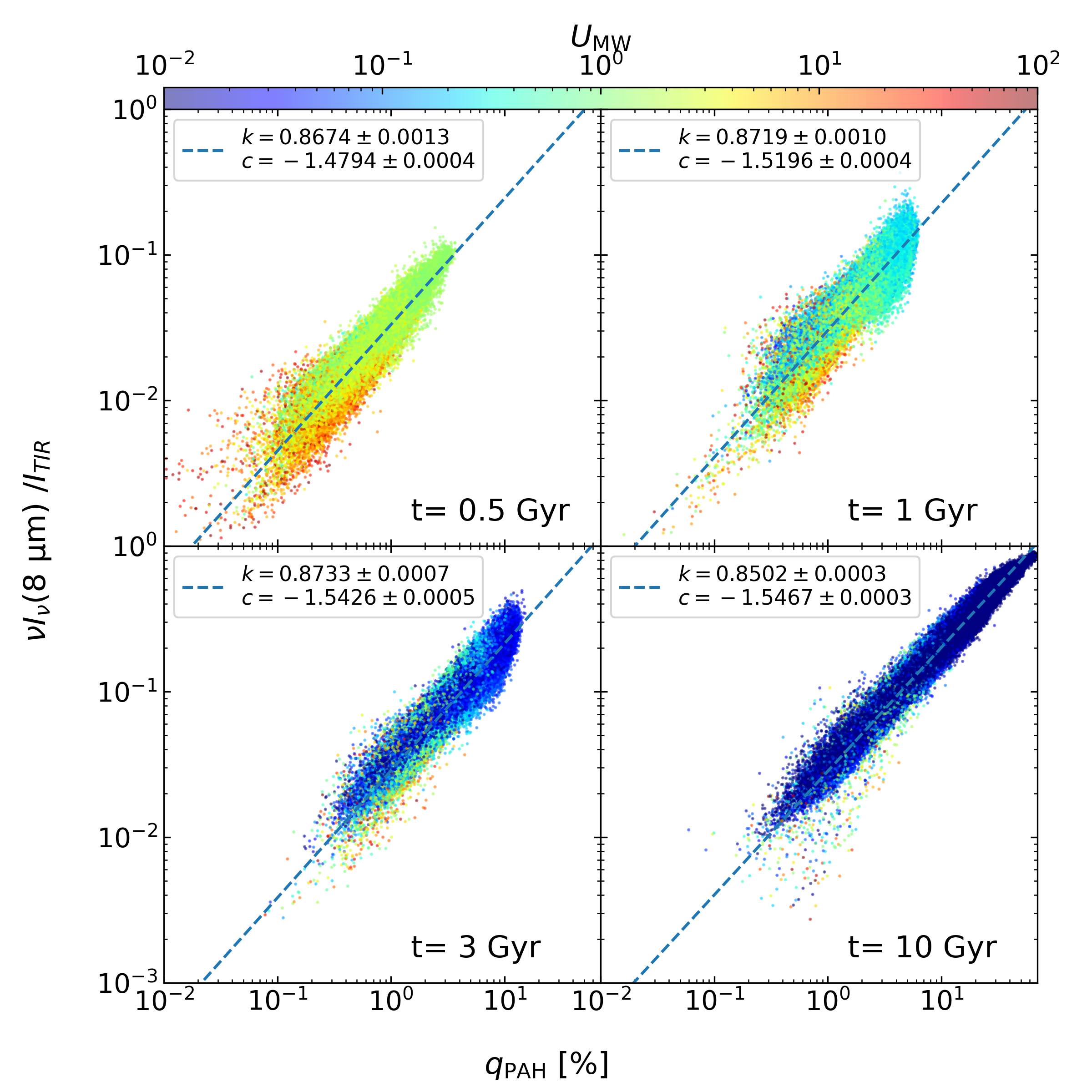}
     \caption{Pixel-based relation between the 8 $\um$-to-total IR intensity ratio and the PAH mass fraction of the Milky Way-like galaxy simulation at $t= 0.5, \ 1.0, \ 3.0,$ and $10$ Gyr from top left to bottom right. The color represents the density-weighted mean intensity at 1000 \AA\ relative to the Milky Way value, as indicated by the color bar on the top. The blue dashed line at each age indicates the best-fit relation of Eq.~(\ref{eq: linear regression}). }%
     \label{fig: 8-TIR ratio PAH mass fraction}
\end{figure}

Here, we use the total infrared intensity estimated by the following equation:
\begin{equation}
    I_\mathrm{TIR} = 0.95 \, \nu I_\mathrm{\nu} (\mathrm{8 \, \um}) + 1.15 \, \nu I_\mathrm{\nu} ( \mathrm{24 \, \um}) + \nu I_\mathrm{\nu} (\mathrm{70 \, \um}) + \nu I_\mathrm{\nu} (\mathrm{160 \, \um}),
    \label{eq: total infrared intensity}
\end{equation}
which is suggested by \citet{Draine&Li2007}. They theoretically indicate that the estimated total infrared intensity differs from the actual one within the relative difference of 10 $\%$ for the radiation fields of $0.1<U_\mathrm{MW}<10^2$.
Here, we expect to maintain higher spatial resolutions of the observed data by utilizing the estimation with bands at up to 160 $\um$ although there are other estimation methods with longer wavelength bands for the total infrared intensity \citep{Boquien2011, Galametz2013}.
Figure \ref{fig: 8-TIR ratio PAH mass fraction} shows the pixel-based relation between PAH mass fraction and $\nu I_\mathrm{\nu} (\mathrm{8 \, \um})/I_\mathrm{TIR}$ at $t= 0.5, \ 1.0, \ 3.0,$ and $10$ Gyr. 
The color represents the density-weighted mean intensity at 1000 \AA\ relative to the Milky Way value, $U_\mathrm{MW}$.
We find that the ratio tightly correlates with $q_\mathrm{PAH}$ and is not affected by the local radiation field of $U_\mathrm{MW}$ unlike $I_\mathrm{\nu} (\mathrm{8 \, \um})/I_\mathrm{\nu} (\mathrm{24 \, \um})$.
Remarkably, the parameters of the best-fit relation (blue) are consistent among the different times within a relative difference of 5\%. This is because both the total and 8 $\um$ intensities are proportional to the strength of the local radiation field \citep{Draine&Li2007}, and hence, the effect of the radiation field is canceled out.
Additionally, we perform linear regression analysis for all of the data including the snapshots at all the ages shown above. We find that the PAH mass fraction of $q_\mathrm{PAH}$ can be estimated using the following equation:
\begin{equation}
    q_\mathrm{PAH} = 66 \, \left(\frac{\mathrm{\nu}I_\mathrm{\nu}(\mathrm{8 \, \um})}{I_\mathrm{TIR}}\right)^{1.2}. \label{eq: qPAH linear regression}
\end{equation}
In Appendix~\ref{Ap: Indicators of the PAH mass fraction of the NGC 628-like galaxy}, we obtain a similar relation for the NGC 628-like galaxy with a deviation of less than a factor of one at $10^{-1} < q_\mathrm{PAH} < 50$. The deviation is greater at lower $q_\mathrm{PAH}$ due to contamination of the continuum of non-aromatic small grains at 8 $\um$.
Therefore, we suggest that $\nu I_\mathrm{\nu} (\mathrm{8 \, \um})/ I_\mathrm{TIR}$ is a good indicator of the PAH mass fraction in the ISM of galaxies in various stages of galaxy evolution as long as the PAH emission is prominent enough. 

\subsubsection{Dependence of the PAH mass fraction on the metallicity and hydrogen surface density}\label{subsubsec: Dependence of the PAH mass fraction on the metallicity and hydrogen surface density}
To demonstrate the usefulness of $\nu I_\mathrm{\nu} (\mathrm{8 \, \um})/ I_\mathrm{TIR}$ as an indicator of the PAH mass fraction, we investigate the spatially resolved relation among the PAH mass fraction, gas-phase metallicity, and hydrogen surface density.
\begin{figure}
        \includegraphics[width=0.5\textwidth]{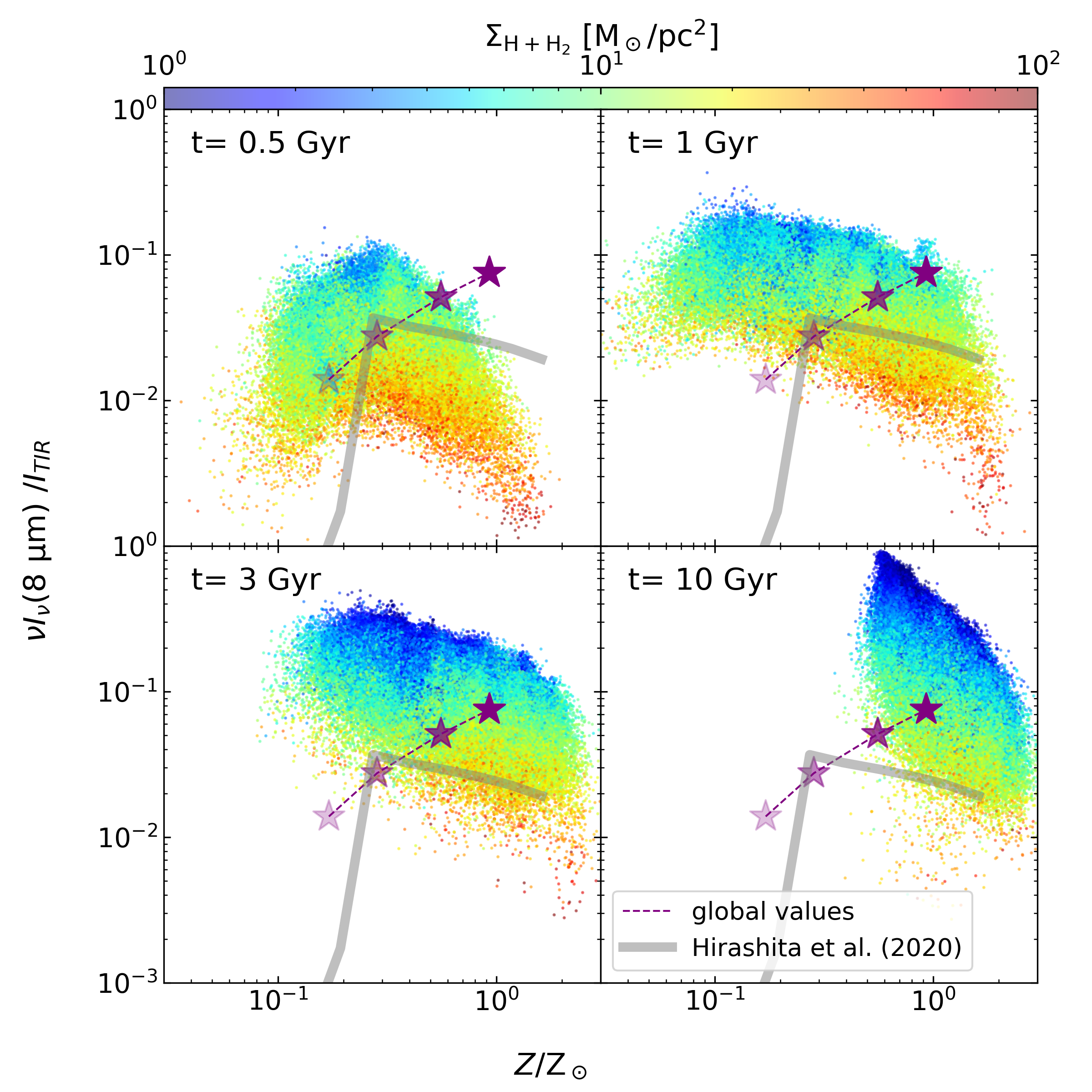}
     \caption{Time evolution of pixel-based relation among the 8 $\um$-to-total IR intensity ratio, gas-phase metallicity, and hydrogen surface density for the Milky Way-like galaxy simulation.
      The panels show the relation at $t= 0.5, \ 1.0, \ 3.0,$ and $10$ Gyr from top left to bottom right. The color bar indicates the hydrogen surface density of each pixel. In each panel, the time evolution of the entire distribution over the four epochs is also indicated by the purple dashed lines. The purple star symbols represent the global intensity ratio and metallicity of the galaxy at $t= 0.5, \ 1.0, \ 3.0,$ and $10$ Gyr from the lower left to the upper right in each panel (the symbol transparency decreases) and are connected by the dashed line. We note that the intensity ratio and metallicity are the intensity- and mass-weighted values, respectively.
      The grey line shows the evolutionary path 
      in the one-zone dust evolution model by \citet{Hirashita&Murga2020} under the condition of $f_\mathrm{dense}=0.5$, $\tau_\mathrm{SF}=5$ Gyr, and $U_\mathrm{MW}=1$. 
     }%
     \label{fig: 8-TIR ratio Metallicity}
\end{figure}
Figure \ref{fig: 8-TIR ratio Metallicity} shows the pixel-based relation between $\nu I_\mathrm{\nu} (\mathrm{8 \, \um})/ I_\mathrm{TIR}$ and metallicity at $t= 0.5, \ 1.0, \ 3.0,$ and $10$ Gyr. The color represents the hydrogen surface density of each pixel.
The purple star symbols represent the global intensity ratio and metallicity of the galaxy at four epochs, and the dashed lines indicate the time evolution of global properties over the epochs (from the lower left to the upper right in each panel).
We find that the global $\nu I_\mathrm{\nu} (\mathrm{8 \, \um})/ I_\mathrm{TIR}$ (PAH mass fraction) increases with time and metallicity. This global trend is also supported by larger samples \citep{Galliano2008b, Remy-Ruyer2015A&A...582A.121R, Shim2023}.
Shattering and accretion happen in separate ISM phases, and their interplay increases small grains (and PAHs) on a galactic scale in our simulations \citep{Hirashita2012}. 
Hence, the global PAH mass fraction evolves with metallicity.

In the pixel-based analysis, the local PAH mass fraction is associated with the local ISM conditions such as metallicity and hydrogen surface density (the projected maps of the hydrogen surface density, metallicity, and PAH mass fraction are shown in Appendix \ref{Ap: Various maps of the Milky Way-like galaxy}).
At $t=0.5$ Gyr, $\nu I_\mathrm{\nu} (\mathrm{8 \, \um})/ I_\mathrm{TIR}$  increases with metallicity in a lower metallicity regime ($Z\lesssim 0.2 \ \Zsun$) while it decreases with metallicity in a higher metallicity regime ($Z\gtrsim 0.2 \ \Zsun$), as suggested by a one-zone dust evolution model \citep[grey line;][]{Hirashita&Murga2020}. We note that the number of pixels in the lower metallicity regime is significantly reduced by the pixel selection
(Section~\ref{subsec: Observational data and analysis}).
At low metallicity, the interplay between accretion and shattering increases the PAH mass fraction, producing a trend of increasing the PAH mass fraction with metallicity.
At high metallicity, dust grows sufficiently through metal accretion, and the effect of coagulation becomes prominent.
Coagulation turns small grains, including PAHs, into larger grains and plays an important role in decreasing the PAH mass fraction at a high-metallicity regime.
At $\geq1.0$ Gyr, $\nu I_\mathrm{\nu} (\mathrm{8 \, \um})/ I_\mathrm{TIR}$ anticorrelates with metallicity since most of the regions reach high metallicity.
We also find that $\nu I_\mathrm{\nu} (\mathrm{8 \, \um})/ I_\mathrm{TIR}$ depends on the hydrogen surface density since coagulation converts small grains, including PAHs, to larger grains in the dense ISM.

These results are roughly consistent with observational studies.
\citet{Egorov2023} report a weak anti-correlation between $R_\mathrm{PAH}$ and oxygen abundance based on the spatially resolved observations of four galaxies with high metallicity
($12+\log\mathrm{(O/H)}\gtrsim 8.4$). \citet{Chastenet2023a} show a tendency to increase $R_\mathrm{PAH}$ with oxygen abundance at low oxygen abundance ($12+\log\mathrm{(O/H)}\lesssim 8.5$) but flatten out at high oxygen abundance ($12+\log\mathrm{(O/H)}\gtrsim 8.5$). However, those observations cover only four galaxies, and future observations are necessary to verify whether there are general trends for the spatially resolved PAH mass fractions to increase and decrease with metallicity at a lower and higher metallicity, respectively.  

Therefore, the spatially resolved relation of the PAH mass fraction against metallicity is related to the efficiencies of coagulation, shattering, and accretion across diverse ISM conditions, and the analysis of the spatially resolved relation leads us to the understanding of dust evolution. 

\section{NGC 628-like galaxy simulation}\label{NGC 628-like galaxy simulation}
NGC 628 is an ideal galaxy for validating our dust model based on spatially resolved data since it is well observed with a high spatial resolution in the PHANGS surveys \citep{Leroy2021, Lee2023}.
Here, we model the NGC 628-like galaxy simulation according to the morphological parameters of NGC 628 (see Appendix \ref{Ap:morphological parameters of NGC628}) to make a fair comparison.
We use the simulation snapshot at $t=10$ Gyr as a representative age of present-day galaxies.
More comprehensive comparisons including other nearby galaxies will be presented in our future work (van der Giessen et al. in preparation).


\subsection{SED comparisons}
\begin{figure}         \includegraphics[width=0.5\textwidth]{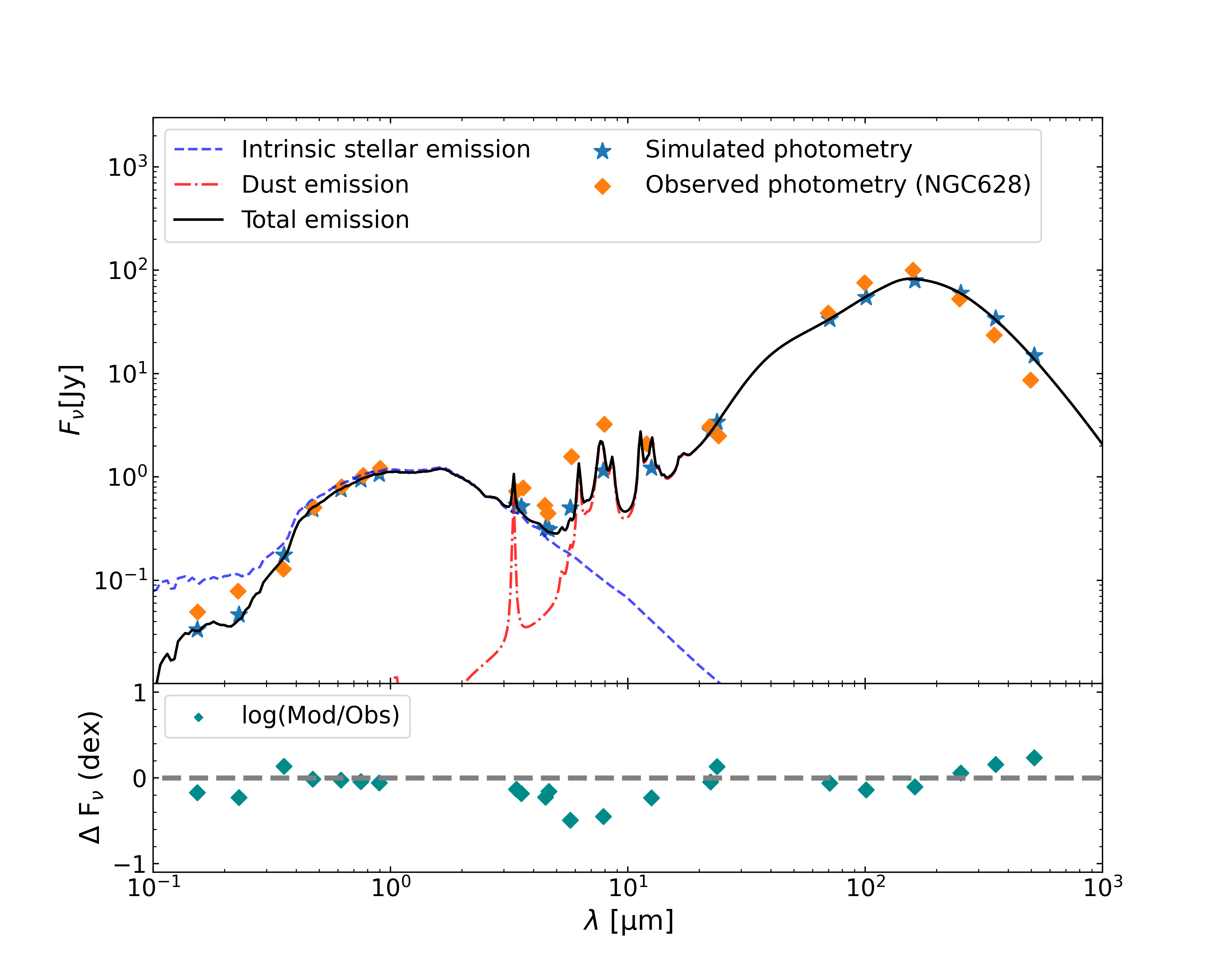}
     \caption{Comparison of the global SED between the NGC 628-like galaxy simulation and actual observations.
     {\it Upper panel}:  Global SED of the NGC 628-like galaxy simulation at $t=10$ Gyr shown with the black solid line.
     Blue dashed and red dash-dotted lines represent the intrinsic stellar and dust emission, respectively. The orange and blue diamonds correspond to the simulated and observed photomeries of the total SEDs (van der Giessen et al. in prep.), respectively.
     {\it Lower panel}: Difference in the photometries between the simulation and observation data.}%
     \label{fig: NGC628 SED comparison}
\end{figure}

Figure \ref{fig: NGC628 SED comparison} shows the total comparison of SED between the simulation and the actual observations of NGC~628 \citep[][blue and orange diamonds, respectively]{Relano2020}.
The total, intrinsic stellar, and dust emission SEDs of the NGC 628-like simulation are also indicated by black solid, blue dashed, and red dash-dotted lines, respectively. The lower panel indicates the difference in the photometries between the simulation and the observations.
Stellar emission at optical and near-IR wavelengths of the simulated galaxy is consistent with that of the observations while the emission at UV wavelengths in the simulation is weaker than that of the observations. This is because the SFR of the simulated galaxy at $t=10$ Gyr ($\dot{M}_* = 0.75$ $\mathrm{M_{\odot}~yr^{-1}}$) is lower than that reported by the observation \citep[$\dot{M}_* = 1.7$ $\mathrm{M_{\odot}~yr^{-1}}$;][]{Leroy2021}. 
On the other hand, dust emission at longer wavelengths of $\lambda \geq20$ $\um$ in the simulation is consistent with that in observations while the PAH emission at MIR wavelengths in the simulation is lower than that in observations by about 0.5 dex.
This result implies that our dust model underestimates the PAH mass, as reported in \citet{Hirashita2020SEDwithPAH} and \citet{Hirashita2023}.
To explore where the PAH mass fraction is underestimated in our simulation, we perform the pixel-based comparison of the PAH mass fraction in the next section.

\subsection{Spatial comparison of PAH mass fraction}\label{subsec: comparison of PAH mass fraction of NGC 628}
\begin{figure*}
    \includegraphics[width=1.0\textwidth]{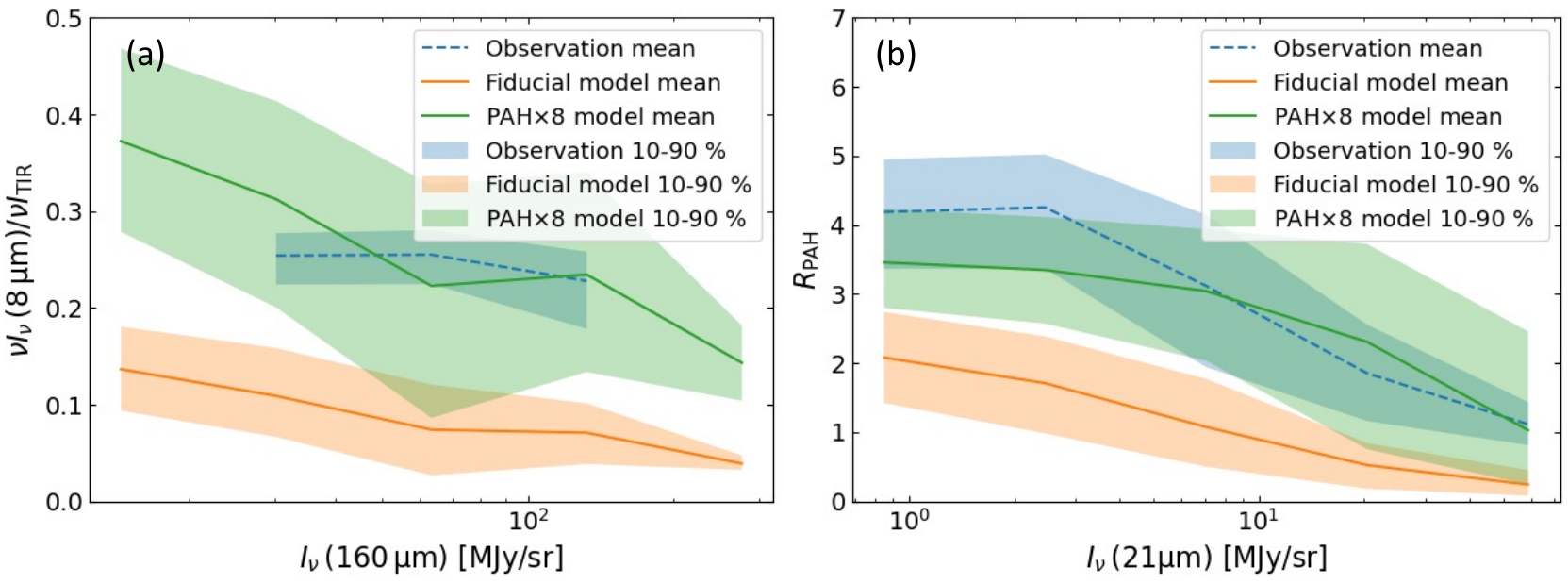}
     \caption{Pixel-based comparison of the indicators of the PAH mass fraction between the NGC 628-like galaxy simulation and actual observations. {\it Panel (a)}: $\nu I_\mathrm{\nu} (\mathrm{8 \, \um})/ I_\mathrm{TIR}$ versus the $\mathrm{160 \, \um}$ intensity for the NGC 628-like galaxy simulation at $t=10$ Gyr (orange) and based on the \textit{Herschel} and \textit{Spitzer} observations for NGC 628 (blue). 
     {\it Panel (b)}: $R_\mathrm{PAH}$ versus the $\mathrm{21 \, \um}$ intensity for the NGC 628-like galaxy simulation at $t=10$ Gyr (orange) and based on the \textit{JWST} observations for NGC 628 (blue).
     In both panels, the green color represents the model, in which the PAH mass is enhanced by a factor of eight. The solid and dashed lines represent the mean value over all pixels of the photometric data of the models and observations, respectively, and the shaded area corresponds to the 10th--90th percentile range. }%
     \label{fig: NGC628 Spitzer Herschel I8TIR raio comparison}
\end{figure*}

We compare the spatially resolved data of $\nu I_\mathrm{\nu} (\mathrm{8 \, \um})/ I_\mathrm{TIR}$ between our simulation and \textit{Herschel}/\textit{Spitzer} observations. The observations cover large scales up to the galactic radius of $r\sim 10$ kpc, but the spatial resolution is about 600 pc. 
Figure~\ref{fig: NGC628 Spitzer Herschel I8TIR raio comparison}(a) shows the comparison of the relation between $\nu I_\mathrm{\nu} (\mathrm{8 \, \um})/ I_\mathrm{TIR}$ and the intensity at $\mathrm{160 \, \um}$. The line and shaded area correspond to the median value and the 10th--90th percentile range, respectively.
Here, the $\mathrm{160 \, \um}$ band traces large grains in radiative equilibrium and represents the dust mass distribution.
We find that $\nu I_\mathrm{\nu} (\mathrm{8 \, \um})/ I_\mathrm{TIR}$ from the simulation is lower
than that of the observation in the entire range of $\mathrm{160 \, \um}$.
This indicates an underestimation of the PAH mass fraction in our model, regardless of location.
To estimate how much additional PAH mass is required in the NGC 628-like galaxy simulation (hereafter, referred to as the fiducial model), we assume eight times the PAH mass in the fiducial model and find that $\nu I_\mathrm{\nu} (\mathrm{8 \, \um})/ I_\mathrm{TIR}$ can be consistent with that of the observation as shown in Figure~\ref{fig: NGC628 Spitzer Herschel I8TIR raio comparison}(a).
Therefore, we conclude that the PAH mass abundance is underestimated by a factor of eight on average in our simulated area.

\begin{figure*}
\includegraphics[width=1.0\textwidth]{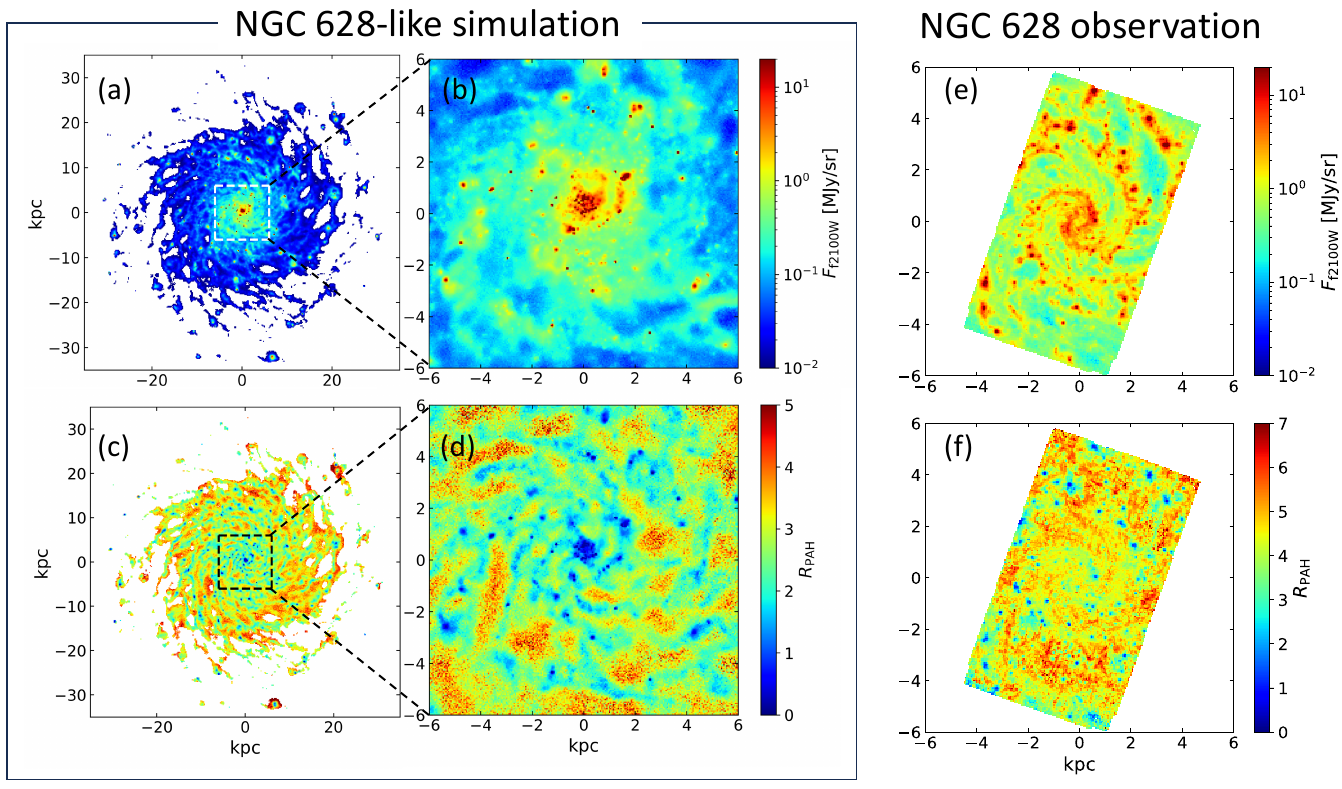}
     \caption{Comparison of $21$ $\um$ intensity and $R_\mathrm{PAH}$ maps between the NGC~628-like galaxy simulation and the \textit{JWST} observations.
     {\it Panels (a) and (b)}: $21$ $\um$ intensity maps of the NGC~628-like galaxy simulation at $10$ Gyr on 100 and 20 kpc scales, respectively. {\it Panels (c) and (d)}: $R_\mathrm{PAH}$ maps of the simulation on 100 and 20 kpc scales, respectively. {\it Panels (e) and (f)}: maps of the $21$ $\um$ intensity and $R_\mathrm{PAH}$ based on the \textit{JWST}/MIRI observations for NGC~628 \citep{Lee2023}, where the spatial resolutions are reduced to 50 pc.}%
     \label{fig: NGC628 I24 RPAH maps}
\end{figure*}
Furthermore, we compare the spatially resolved data from our simulation and the \textit{JWST} observations \citep{Lee2023}, which have 12 times higher spatial resolution than the \textit{Herschel} observations (a pixel size $\sim$ 50 pc).
Here, instead of $I_\mathrm{\nu} (\mathrm{8 \, \um})/I_\mathrm{\nu} (\mathrm{24 \, \um})$, we use the intensity ratio of $R_\mathrm{PAH}$ (Eq.\ (\ref{eq: R_PAH})).
Fig.~\ref{fig: NGC628 Spitzer Herschel I8TIR raio comparison}(b) shows the comparison of $R_\mathrm{PAH}$ to the intensity at $\mathrm{21 \, \um}$ between our simulation (orange) and the observation (blue).
We find that, in both simulation and observation, $R_\mathrm{PAH}$ decreases as the $\mathrm{21 \, \um}$ intensity increases.
This trend can be explained by our model in which coagulation is more efficient in denser regions with higher $\mathrm{21 \, \um}$ intensities. 
Figures \ref{fig: NGC628 I24 RPAH maps}(a)/(c) and (b)/(d) show the intensity/$R_\mathrm{PAH}$ maps at $21$ $\um$ of the entire NGC 628-like galaxy at $t=10$ Gyr and the zoomed central 6 kpc region, respectively.
In Fig. \ref{fig: NGC628 I24 RPAH maps}(c), $R_\mathrm{PAH}$ gradually increases toward the outer regions of the galaxy, since shattering is more dominant in the diffuse ISM than coagulation. 
In the central region of the galaxy (Fig. \ref{fig: NGC628 I24 RPAH maps}(d)), $R_\mathrm{PAH}$ is lower in the spiral arms and clumps, where the intensity at $21$ $\um$ is higher. This is because coagulation is dominant in the dense ISM.
These trends are consistent with the results revealed by recent \textit{JWST} observations. Figs.~\ref{fig: NGC628 I24 RPAH maps}(e) and (f) show the observed maps of $21$ $\um$ intensity and $R_\mathrm{PAH}$ maps, respectively \citep{Lee2023}.
These observations also indicate that $R_\mathrm{PAH}$ is lower in the spiral arms and clumps, where the $21$ $\um$ intensity is higher.
Alternatively, it is also possible that PAHs are destroyed by UV radiation from young stars in star-forming regions in the dense ISM, as discussed in the literature \citep[e.g.,][]{Egorov2023, Chastenet2023b} since $\mathrm{21 \, \um}$ intensity is expected to increase as UV radiation field becomes stronger.
Thus, to discriminate these scenarios, we have to implement PAH destruction due to UV radiation from young stars in future simulations \citep[e.g.,][]{Nanni2023}.
We also find that the value of $R_\mathrm{PAH}$ from the simulation is lower than that from the observation, as discussed above: 
$R_\mathrm{PAH}$ of the model with the PAH mass enhanced by a factor of eight is consistent with that of the observations, implying that further enhancement of small dust grains in our model is required.

\subsection{Insights into additional PAH formation mechanisms} 
As discussed above, our dust evolution model underestimates the PAH mass fraction in the entire galaxy, and thus, additional mechanisms to enhance the PAH production in our model are required.
In our model, we assume that the main driver of PAH formation is the shattering of large carbonaceous grains, but PAH formation in the envelopes of carbon-rich AGB stars may also be considered. Furthermore, the shape of the grain size distribution for the NGC 628 galaxy at $t=10$ Gyr is determined by the balance between coagulation and shattering, and hence, PAHs might be lost too efficiently by coagulation in the dense ISM. Here, we discuss the possible scenarios to enhance PAHs in the ISM of our simulations. 

\citet{Cherchneff1992} suggest that the stellar envelopes of carbon-rich AGB stars can be a site of the PAH formation.
Indeed, many observational studies reveal evidence of the PAH formation in the stellar envelopes of AGB stars by identifying the PAH features in the spectra of the individual stars in the Magellanic Clouds and our Galaxy \citep{Sloan2007,Sloan2014, Matsuura2014}. 
However, some studies based on the Magellanic Clouds report that the presence of PAHs in the ISM can not be explained entirely only by carbon-rich AGB stars \citep{Matsuura2013, Sandstrom2010, Chastenet2019}. 
Moreover, because of the higher metallicity in NGC 628 than in the Magellanic Clouds, dust is likely produced more efficiently by accretion than by stellar dust production.
Thus, it is not probable that stellar PAH production is dominant over the PAH formation by the shattering of large grains (predominantly formed by accretion) in NGC 628.


In our model, coagulation in the dense ISM efficiently depletes PAHs.
\citet{Hirashita2023} constructs a hypothetical model in which small carbonaceous grains are not involved in some interstellar processing, especially coagulation, and succeeds in enhancing the PAH abundance to a level consistent with the observed intensity of PAH emission in the Milky Way.
There are some possible mechanisms that suppress the coagulation of small grains.
\citet{Akimkin2023} suggest that collisions between charged small grains are inhibited by Coulomb repulsion, whereas collisions between large grains are energetic enough to overcome the Coulomb barrier.
This study indicates that a key to the PAH deficiency problem in our model lies in the treatment of grain charges.
Alternatively, the interaction of charged grains with magnetic fields may lead to grain velocities that could cause shattering even in the dense ISM \citep{Yan2004}. If PAH formation by shattering occurs in the dense star-forming regions, PAHs could be efficiently illuminated by UV.
PAH formation may also be activated through rotational disruption caused by radiative torques in such illuminated regions \citep{Hoang2019, Hoang2019Nat}.
These additional PAH formation mechanisms in irradiated regions could also contribute to resolving the underestimate of PAH emission.
These arguments imply that some processes not involved in our simulations may provide keys to resolving the underestimate of the PAH abundance.

\section{Conclusion}\label{Conclusion}
In this paper, we investigate how dust evolution is reflected on IR spectral features and we provide observational signatures of the grain size evolution in ISM, in particular, by exploring indicators of the PAH mass fraction. We modeled the dust evolution for two simulations representing star-forming galaxies similar to the Milky Way and NGC 628 using GADGET4-OSAKA \citep{Romano2022Dust}. We simulated the evolution of grain size distributions in a manner consistent with the physical conditions of the ISM (see Section \ref{subsec: Isolated galaxy simulations}) by taking into account interstellar dust processes, including stellar dust production, SN destruction, shattering, coagulation, and metal accretion.


Furthermore, we separated the dust compositions into silicate and carbonaceous dust grains, and PAHs according to the abundance of Si and C and the mass fraction of the dense gas-phase in each gas particle in post-processing.
Then, we performed radiative transfer calculations with SKIRT under the dust optical properties consistent with the grain size distribution and grain composition of the hydrodynamic simulations (see Section \ref{subsec:SKIRT}) to investigate the evolution of SED and emission from the galaxies and indicators of PAH mass fraction.
Finally, we compared the indicators of the PAH mass fraction between the NGC 628-like simulation and the observations.

We summarize our results as follows:
\begin{enumerate}
    \item
    We show the history of dust evolution and star formation
    for the Milky Way-like and NGC 628-like galaxy simulations (see Fig. \ref{fig: Dust evolutions of isolated galaxy models}). The Milky Way-like galaxy shows more rapid dust formation because of the earlier starburst and metal production than the NGC 628-like galaxy. However, the Milky Way-like galaxy shows less PAH formation in the ISM since it contains a larger amount of dense gas, where shattering is inefficient.
    These results suggest that different ISM conditions and SFHs cause different dust evolution and PAH mass fractions among galaxies.
    \item 
    We investigate the detailed time evolution of observable properties for the Milky Way-like simulation. In the early phase, only the emission of large dust grains can be observed around $\lambda=70$ $\um$ in the SED, as dust production is dominated by stellar sources (see Fig.~\ref{fig: SED evolution of the MW-like galaxy}). Afterward, the emission of small grains and PAHs becomes noticeable around $\lambda=10$ $\um$, since shattering acts as a source of small grains in diffuse regions. In addition, accretion also increases the abundance of small grains (see panels with $t=0.5$ Gyr in Fig.~\ref{fig: Evolution of band-maps of the MW-like galaxy simulation}). 
    Finally, the functional shape of the grain size distribution in each region of the ISM converges to that determined by the balance between shattering and coagulation rate according to the local physical condition. Consequently, the emission from large and small grains is relatively prominent in the inner and outer regions of the galaxy, respectively (see panels with $t=10$ Gyr in Fig.~\ref{fig: Evolution of band-maps of the MW-like galaxy simulation}).
    \item 
    We find that $I_\mathrm{\nu} (\mathrm{8 \, \um})/I_\mathrm{\nu} (\mathrm{24 \, \um})$ is correlated with the PAH mass fraction, but the local radiation field influences it (see Fig.~\ref{fig: 8-24 ratio PAH mass fraction}). The 8 $\um$ band is almost purely dominated by PAH band emission, while the 24 $\um$ band is contaminated by the emission of large grains; as a result, $I_\mathrm{\nu} (\mathrm{8 \, \um})/I_\mathrm{\nu} (\mathrm{24 \, \um})$ increases with the PAH mass fraction. However, stronger radiation fields cause higher dust temperatures, shifting the large grain emission toward shorter wavelengths and raising the 24 $\um$ emission. Consequently, $I_\mathrm{\nu} (\mathrm{8 \, \um})/I_\mathrm{\nu} (\mathrm{24 \, \um})$ decreases in regions with strong radiation. Hence, $I_\mathrm{\nu} (\mathrm{8 \, \um})/I_\mathrm{\nu} (\mathrm{24 \, \um})$ can be used as an indicator of the PAH mass fraction with a tendency of underestimating PAH mass fraction in regions with strong radiation. 
    \item 
    We find that $\nu I_\mathrm{\nu} (\mathrm{8 \, \um})/ I_\mathrm{TIR}$ is tightly correlated with the PAH mass fraction and is not affected by the local radiation field (see Fig.~\ref{fig: 8-TIR ratio PAH mass fraction} and Eq. \ref{eq: qPAH linear regression}). This is because both the total and 8 $\um$ intensities are proportional to the strength of the local radiation field; hence, the effect of the radiation field is canceled out. Therefore, we suggest that $\nu I_\mathrm{\nu} (\mathrm{8 \, \um})/ I_\mathrm{TIR}$ is a good indicator of the PAH mass fraction in various stages of galaxy evolution. 
    \item
    We find that the average PAH mass fraction or $\nu I_\mathrm{\nu} (\mathrm{8 \, \um})/ I_\mathrm{TIR}$ in the entire galaxy evolves with metallicity (see Fig. \ref{fig: 8-TIR ratio Metallicity}). This is because the interplay between shattering and accretion produces PAHs in the entire galaxy. On the contrary, based on the pixel-based analysis, local values of $\nu I_\mathrm{\nu} (\mathrm{8 \, \um})/ I_\mathrm{TIR}$ decrease with metallicity in the ISM at higher metallicity ($Z>0.2 \ \Zsun$), while it increases at lower metallicity ($Z\leq0.2 \ \Zsun$). At low metallicity, metal enrichment causes more efficient shattering, which tends to increase the PAHs. At high metallicity, coagulation is rather important in depleting PAHs.
    We also find that $\nu I_\mathrm{\nu} (\mathrm{8 \, \um})/ I_\mathrm{TIR}$ depends on the gas surface density since coagulation converts PAHs to large grains more efficiently in denser environments.
    Therefore, the spatially resolved PAH mass fraction helps us interpret the efficiency of accretion, coagulation, and shattering in various ISM phases.
    \item 
    We compared the total SED in the NGC 628-like simulation with the observed one (see Fig.~\ref{fig: NGC628 SED comparison}). Although dust emission at longer wavelengths of $\lambda \geq20$ $\um$ in the simulation is consistent with that in the observations, the PAH emission in the MIR wavelengths in the simulation is lower than that in observations by about 0.5 dex. This result implies that our dust model underestimates PAH masses.
    \item 
    We compared the spatially resolved data of $\nu I_\mathrm{\nu} (\mathrm{8 \, \um})/ I_\mathrm{TIR}$ in the NGC 628-like galaxy simulation and the \textit{Herschel}/\textit{Spitzer} data (see Fig.~\ref{fig: NGC628 Spitzer Herschel I8TIR raio comparison}(a)).  We find that the values of $\nu I_\mathrm{\nu} (\mathrm{8 \, \um})/ I_\mathrm{TIR}$ from the simulation are lower than those from the observations, implying that eight times more PAH mass is required in our simulation.
    \item 
    We also compared the spatially resolved data from our simulation with the \textit{JWST} observations with high spatial resolution (see Fig.~\ref{fig: NGC628 Spitzer Herschel I8TIR raio comparison}(b)). We find that $R_\mathrm{PAH}$, a PAH indicator composed of \textit{JWST} bands (Eq. \ref{eq: R_PAH}), decreases with the $\mathrm{21 \, \um}$ intensity in both the simulation and observations. This trend can be explained by the fact that coagulation is more efficient in a denser environment. This comparison also shows that the values of $R_\mathrm{PAH}$ from the simulation are lower than those from the observations.
    In our model, coagulation may deplete PAHs too efficiently in dense ISM. Therefore, it may be required to prevent small carbonaceous grains from being lost through coagulation in the ISM.
\end{enumerate}
These results suggest that spatially resolved comparison between observations and theoretical models using observable signatures of grain size evolution (e.g., $\nu I_\mathrm{\nu} (\mathrm{8 \, \um})/ I_\mathrm{TIR}$, $I_\mathrm{\nu} (\mathrm{8 \, \um})/I_\mathrm{\nu} (\mathrm{24 \, \um})$, $R_\mathrm{PAH}$) leads us to a quantitative investigation of dust evolution in various ISM phases.
In particular, future JWST observations for a large sample of galaxies with high spatial resolution will systematically reveal the efficiency of dust processes in the diffuse and dense ISM.

\section*{Acknowledgements}
KM is a Ph.D. fellow of the Flemish Fund for Scientific Research (FWO-Vlaanderen) and acknowledges the financial support provided through Grant number 1169822N.
Numerical computations were performed on the Cray XC50 at the Center for Computational Astrophysics, National Astronomical Observatory of Japan, and on the SQUID at the Cybermedia Center, Osaka University as part of the HPCI system Research Project (hp220044,hp230089). 
This work is supported in part by the MEXT/JSPS KAKENHI grant numbers 20H00180 and 22K21349 (KN), and 21J20930, 22KJ2072 (YO). 
KN acknowledges the support from the Kavli IPMU, World Premier Research Center Initiative (WPI), where a part of this work was conducted. 
HH thanks the National Science and Technology Council for support through the grant 111-2112-M-001-038-MY3 and the Academia Sinica for Investigator Award AS-IA-109-M02. MR acknowledges the support from the project PID2020-114414GB-100, financed by MCIN/AEI/10.13039/501100011033. SvdG and IDL acknowledge financial support from the European Research Council (ERC) under de European Union's Horizon 2020 research and innovation program DustOrigin (ERC-2019-StG-851622) and by the Flemish Fund for Scientific Research (FWO-Vlaanderen) through the research project G023821N.
We thank Prof. Nakagawa for many fruitful discussions.

\bibliography{dust_paper}    

\begin{appendix} 

\section{Initial conditions of the NGC 628-like galaxy}\label{Ap:morphological parameters of NGC628}
We constructed the initial condition for the NGC 628-like simulation using the DICE code \citep{Valentin2016Dice}, which computes the gravitational potential on a multilevel Cartesian mesh and considers the dynamical equilibrium of multiple material components. The process is detailed in the following.
We build up the structural properties of the five components: the dark matter halo, stellar disk and bulge, and gaseous disk and halo.
First, the total halo is characterized by a viral mass of $1.2\times10^{12}$ $\Msun$, a virial velocity of $v_{200}=150~\mathrm{km~s^{-1}}$, and the spin parameter $ \lambda= 0.04$. These parameters are determined so that the rotation curve of the simulation is consistent with that of the NGC 628 observations \citep{Aniyan2018}.
The dark matter halo with a mass of $1.0\times10^{12}$ $\Msun$ follows the Navarro–Frenk–White \citep{Navarro1997} profile with a concentration parameter of $c= 25$ and an effective radius of 100 kpc.
The gaseous halo follows the same profile but has 1 \% of the total mass, $1.2\times10^{10}$ $\Msun$.
Second, we obtain the morphological parameters of the gaseous and stellar disk by fitting the observationally estimated radial profiles of the gas and stellar surface densities (S. van der Giessen et al. in prep.) with the Sersic profile given by,
\begin{equation}
    \Sigma(r) = \Sigma_{0} \exp{-b_n\Bigg[\Big(\frac{r}{R_e}\Big)^{1/n}-1\Bigg]},
    \label{eq: Sersic equation}
\end{equation}
where $r$, $R_e$, $ \Sigma_{0}$, $n$, and $b_n$ are the radius in the cylindrical coordinate, the half-mass radius, the surface density at the half-light radius, S\'{e}rsic index, and a coefficient describing the steepness of the exponential decline, respectively.
The parameters obtained in the fitting are summarized in Table~\ref{table: NGC628 parameters of the Sersic profiles}.
Furthermore, we assume that the vertical density profile of the galaxy is described by
\begin{equation}
    \rho(z) \propto \exp (-z/h_z),
    \label{eq: Sersic equation height}
\end{equation}
where $z$ and $h_z$ are the height in the cylindrical coordinate and the scale height, respectively. The scale height was obtained from \citet{Aniyan2018}.
The gaseous and stellar disks have masses of $2.6\times10^{10}$ and $3.5\times10^{10}$ $\Msun$, respectively.
Finally, the stellar bulge follows the Einasto profile \citep{Einasto1965} with a structural parameter of $\alpha=1.0$ and a mass of $9.26\times10^{9}$ $\Msun$.
\begin{table}
\caption{Morphological parameters of the gas and stellar components in the NGC 628-like galaxy simulation}             
\label{table: NGC628 parameters of the Sersic profiles}      
\centering                          
\begin{tabular}{l | c  c c c c}        
\hline\hline                 
Type & $\Sigma_0$ $(\mathrm{\Msun/pc^{-2}})$ & $R_e$ $(\mathrm{kpc})$ & $n$ & $b$ & $h_z$ $(\mathrm{kpc})$\\    
\hline                        
Star & 12 & 12 & 1.5 & 3.9 & 0.4 \\ 
Gas & 15 & 6.5 & 0.5 & 0.26 & 0.4 \\ 
\hline                                   
\end{tabular}
\end{table}
\begin{figure}[h] 
        \includegraphics[width=0.48\textwidth]{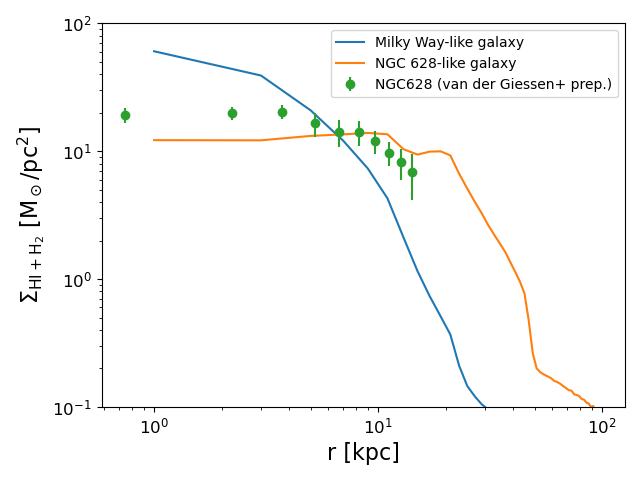}
     \caption{Radial profiles of the hydrogen gas surface densities of NGC 628-like and Milky Way-like galaxies at $t=0$ Gyr (orange and blue lines, respectively), compared with that from the actual observations of NGC 628 (green circle; van der Giessen et al. in prep.).}%
     \label{fig:radial profiles of NGC628 and MW}
\end{figure}

Figure~\ref{fig:radial profiles of NGC628 and MW} shows the comparison of the radial profiles of the hydrogen gas surface densities between NGC 628-like and Milky Way-like galaxies at $t=0$ Gyr (blue and orange lines, respectively).
For reference, the radial profile from the actual observations for NGC 628 (van der Giessen et al. in prep.) is also indicated by the green circle.
Under the initial conditions, the gas radial profile of the NGC 628-like galaxy is flatter than that of the Milky Way-like galaxy, and the gas density in the central region of the NGC 628-like galaxy is not as high.
Thus, the star formation of the NGC 628-like galaxy occurs slowly compared to that of the Milky Way-like galaxy (see Section~\ref{sec: Dust evolution in isolated galaxy simulations}).


\section{Comparison of $\nu I_\nu (\mathrm{8 \, \um})/ I_\mathrm{TIR}$ between the Milky Way-like and NGC 628-like galaxies}\label{Ap: Indicators of the PAH mass fraction of the NGC 628-like galaxy}

\begin{figure}
        \includegraphics[width=0.5\textwidth]{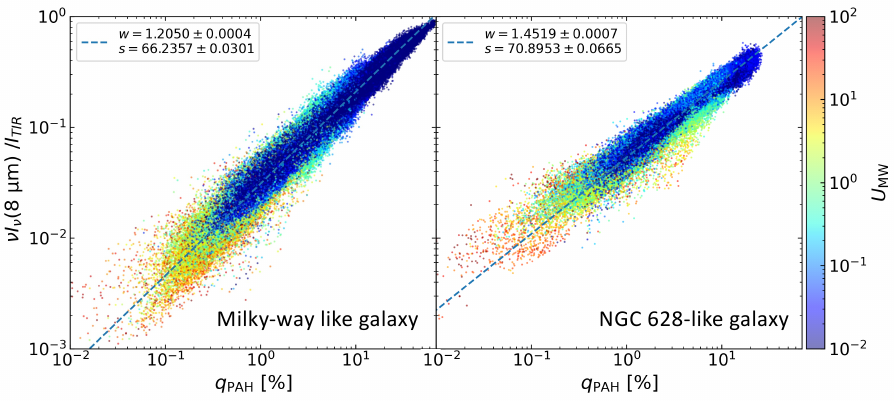}
     \caption{Pixel-based relation between the 8 $\um$-to-total IR intensity ratio and the PAH mass fraction of the Milky Way-like and NGC 628-like galaxies using all snapshots at $t= 0.5, \ 1.0, \ 3.0,$ and $10$ Gyr (left and right panels, respectively). The color represents the density-weighted mean intensity at 1000 \AA\ relative to the Milky Way value as indicated by the color bar. Blue dashed lines indicate the best-fit relation of Eq.~(\ref{eq: qPAH linear regression of different galaxies}), and the legend indicates the best-fit parameter. }%
     \label{fig: 8-TIR ratio PAH mass fraction of MW and NGC628}
\end{figure}
To discuss the robustness of the indicator of the PAH mass fraction, $\nu I_\nu (\mathrm{8 \, \um})/ I_\mathrm{TIR}$, in different galaxies, we compare the pixel-based relation for the Milky Way-like and NGC 628-like galaxies, as shown in Fig.~\ref{fig: 8-TIR ratio PAH mass fraction of MW and NGC628}.
The blue dashed lines in each panel show the linear regression for the relation between $\nu I_\nu (\mathrm{8 \, \um})/ I_\mathrm{TIR}$ and the PAH mass fraction, which is formulated by the following equation:
\begin{equation}
    q_\mathrm{PAH} = s \, \left(\frac{\mathrm{\nu}I_\mathrm{\nu}(\mathrm{8 \, \um})}{I_\mathrm{TIR}}\right)^{w}. \label{eq: qPAH linear regression of different galaxies}
\end{equation}
We find that the fitting relation between the Milky Way-like and NGC 628-like galaxies is slightly different.
Relative to the relation of the Milky Way-like galaxy, that of the NGC 628-like galaxy has a $70$, $50$, and $30$ $\%$ difference at $q_\mathrm{PAH}=0.1$, $1$, and $10$, respectively.
This is because the 8 $\um$ band is more contaminated by the continuum of non-aromatic small grains at lower $q_\mathrm{PAH}$, causing the difference of significance of the PAH band emission against the continuum level at 8 $\um$ between the two galaxies
Therefore, the relation between $\nu I_\mathrm{\nu} (\mathrm{8 \, \um})/ I_\mathrm{TIR}$ and $q_\mathrm{PAH}$ can vary between galaxies, but the variance might be small. 

\section{Various maps of the Milky Way-like galaxy}\label{Ap: Various maps of the Milky Way-like galaxy}
\begin{figure*}        
\includegraphics[width=1.0\textwidth]{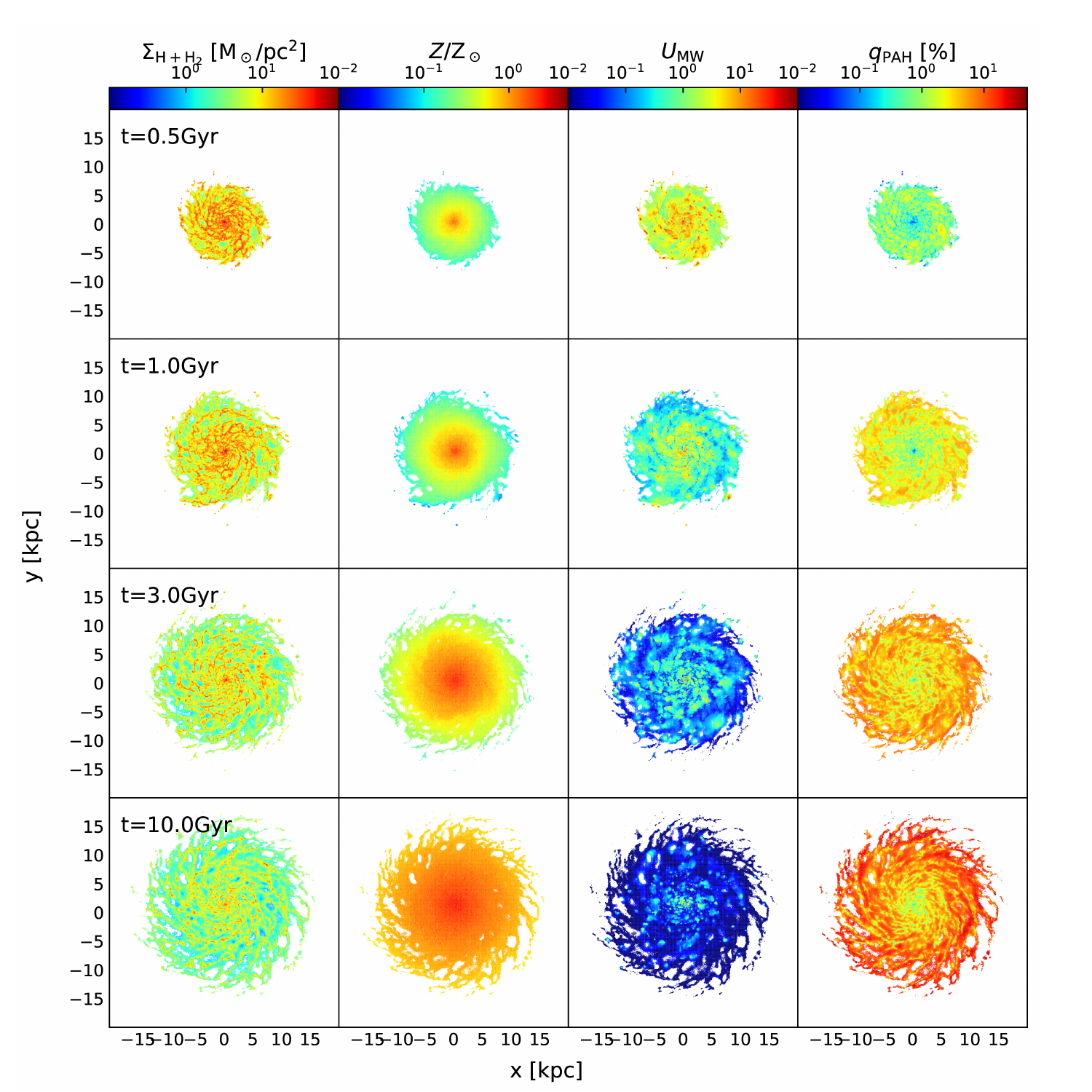}
     \caption{Maps of the hydrogen surface density, metallicity, local radiation field, and PAH mass fraction (left to right) of the Milky Way-like galaxy simulation at $t= 0.5, \ 1.0, \ 3.0,$ and $10$ Gyr (top to bottom).}%
     \label{fig: Appendix physical maps of the MW-like galaxy simulation}
\end{figure*}

Figure \ref{fig: Appendix physical maps of the MW-like galaxy simulation} shows the maps of the hydrogen surface density, metallicity, local radiation field, and PAH mass fraction, $q_\mathrm{PAH}$, (left to right columns) of the Milky Way-like galaxy simulation at $t= 0.5, \ 1.0, \ 3.0,$ and $10$ Gyr (top to bottom panels). The metallicity and local radiation field are normalized by the Milky Way values. 
We only show the pixels that satisfy $I_\nu>0.01$ $\mathrm{MJy\,sr^{-1}}$ at all the \textit{Spitzer} and \textit{Herschel} bands; that is,
the pixels used for the analysis (Section \ref{subsec:SKIRT}).
The maps of the hydrogen surface density show spiral arms and clumps inside the arms, which reflect the resulting complex structures in the ISM containing diffuse and dense regions. 
In contrast, the metallicity maps show a smoother axisymmetric gradient in the radial direction.
The metallicity reaches solar at $t=10$ Gyr in the entire disk.
In the maps of the local radiation field, the star-formation regions associated with stellar clusters along the spiral arms are prominent (redder colors). Given the monotonic decrease in the SFR of the Milky Way-like galaxy, the star-forming regions shrink with time in terms of spatial distributions and sizes.
The PAH mass fraction, $q_\mathrm{PAH}$, in the entire galaxy increases with time. PAHs are relatively deficient in dense regions, especially in spiral arms and the galaxy center, because of inefficient PAH formation by shattering and efficient PAH depletion by coagulation.

     \label{fig: Appendix NGC628 I24 RPAH maps}
\end{appendix}

\end{document}